\documentclass[journal,article,submit,moreauthors,pdftex,10pt,a4paper]{mdpi}

\firstpage{1}
\articlenumber{x}
\doinum{10.3390/------}
\pubvolume{xx}
\pubyear{2017}
\copyrightyear{2017}
\externaleditor{Academic Editor: name}
\history{Received: date; Accepted: date; Published: date}

\Title{Plasmonic physics of 2D crystalline materials}

\Author{Zahra Torbatian $^1$  and  Reza Asgari $^{1,2}$*}

\address{%
$^{1}$ School of Nano Science, Institute for Research in Fundamental Sciences (IPM), Tehran 19395-5531, Iran ; z.torbatian@ipm.ir\\
$^{2}$ School of Physics, Institute for Research in Fundamental Sciences (IPM), Tehran 19395-5531, Iran ; asgari@theory.ipm.ac.ir}

\corres{Correspondence: asgari@ipm.ir; Tel.: +98-212-283-5061}

\abstract{
Collective modes of doped two-dimensional crystalline materials, namely graphene, MoS$_2$ and phosphorene, both monolayer and bilayer structures, are explored using the density functional theory simulations together with the random phase approximation. The many-body dielectric functions of the materials are calculated using an {\it ab initio} based model involving material-realistic physical properties. Having calculated the electron energy-loss, we calculate the collective modes of each material considering the in-phase and out-of-phase modes for bilayer structures. Furthermore, owing to many band structures and intreband transitions, we also find high-energy excitations in the systems. We explain that the material-specific dielectric function considering the polarizability of the crystalline material such as MoS$_2$ are needed to obtain realistic plasmon dispersions. For each material studied here, we find different collective modes and describe their physical origins.
}


\begin{document}

\section{Introduction}

In a groundbreaking theoretical concept in the early 1950s, Bohm and Pines \cite{bohm} proved that excitations in long-range Coulomb interacting systems can be decomposed into two separate sectors, namely the high-energy collective excitations, so-called longitudinal bulk plasmons, and the low-energy single-electron excitations. Well-defined plasmon oscillations exist if the momentum of carriers in the system is much smaller than the Thomas--Fermi wavevector. Moreover, the plasmon branch enters the single-electron excitation region where at this point, the collective energy of the plasmon dissipates into single electron-hole excitations. This process is known as the Landau damping~\cite{landau}. The Bohm--Pines result is consistent with the classical plasma picture and was the first demonstration of the idea of the renormalization group theory in physics.

Plasmonics is based on interaction processes between electromagnetic radiation and itinerant charges (electrons or holes) at metallic or doped semiconductor interfaces or in small metallic nanostructures.~Although it is well known that there are two main ingredients of plasmonics, namely~surface plasmon polaritons and localized surface plasmons, it is often far from trivial to appreciate the interlinked nature of many of the phenomena and applications of this field. This is compounded by the fact that throughout the 20th Century, surface plasmon polaritons have been rediscovered in a variety of different contexts. Accordingly, the science of plasmonics is dealing with generation, manipulation and detection of surface plasmon polaritons.

Surface plasmon-polaritons are electromagnetic surface waves
coupled to plasmon modes of the itinerant charges (electrons or holes), propagating
along the interface between a dielectric and a~conductor. Therefore, surface plasmon-polaritons are bound modes whose fields decay exponentially away from the interface, and therefore, plasmonics opened the possibility for manipulating light and controlling light-matter interactions at scales below the diffraction limit.

Propagation of electromagnetic waves at the interfaces of plasmas with other dielectrics depends strongly on the
interface geometry. A surface plasmon was predicted by Ritchie in 1957 \cite{ritchie}. The~plasmon mode shows dispersions of
the various plasmon modes at the metal-vacuum surface. The~two dispersion branches of the surface plasmon, the dispersion-less
branch of the longitudinal bulk plasmon ($\omega_p$) and the dispersion of the so-called multipole surface plasmon, can be found
along with a~linear dispersion for an electromagnetic wave, $\omega=ck$ where $c$ is the speed of light in vacuum, parallel to its
surface stemming from the coupling of a photon and a plasmon at the interface. The coupling between the light and the longitudinal
bulk plasmon leads to a splitting of the $(\omega-k)$ dispersion curves for the excitations, which form a photon dispersion
and the bulk plasma mode as the joint of the photon mode and surface plasmon mode. For small wave vectors, dispersion of the surface
plasmon mode asymptotically approaches the light-line. For the large wavevector, in the local approximation, this surface plasmon
approaches asymptotically a constant frequency, which for metals is $\omega = \omega_p/ \sqrt{2}$.

Let us focus on the interaction of metals with electromagnetic fields using a classical approach based on Maxwell's equations. Small sizes of metallic nanostructures on the order of a few nanometers could be qualitatively described by semiclassical mechanics. The reason for that is the high density of free carriers results in minute spacings of the electron energy levels compared to thermal excitations of energy $k_BT$ at room temperature. Moreover, the optical response of metals clearly depends on the frequency, and therefore, we have to take into account the non-locality in time and space by generalizing the linear relationships to:
\begin{eqnarray}
&&{\bf D}({\bf r},t)=\epsilon_0 \int dt' d{\bf r}' \varepsilon({\bf r}-{\bf r}', t-t') {\bf E}({\bf r}', t') \nonumber\\
&&{\bf J}({\bf r},t)=\int dt' d{\bf r}' \sigma({\bf r}-{\bf r}', t-t') {\bf E}({\bf r}', t')
\end{eqnarray}

It should be noticed that we have implicitly assumed that all length scales are significantly larger than the lattice spacing of the material, i.e., the impulse response functions do not depend on absolute spatial and temporal coordinates, but only on their differences. A fundamental relationship between the dielectric function and the conductivity is given by:

\begin{equation}
\varepsilon({\bf k}, \omega)=1+\frac{i \sigma({\bf k}, \omega)}{\epsilon_0 \omega}
\end{equation}

The general form of the dielectric response $\varepsilon({\bf k}, \omega)$, in the interaction between the light and metals, can be simplified to the limit of a spatially local response through $\varepsilon({\bf k}, \omega) = \varepsilon(\omega)$. The simplification is valid as long as the wavelength in the material is significantly longer than all characteristic dimensions such as the size of the unit cell or the mean free path of the electrons. In general, $\varepsilon({\bf k}, \omega)$ is a complex valued function, and the imaginary part of the dielectric function determines the amount of absorption inside the medium. Most importantly, including quantum interlayer contributions leads to increasing the imaginary part of the dielectric function, and it turns out that the effects of quantum mechanics are very vital in systems in which interlayer transitions play an important role.

It is basically known that the traveling-wave solutions of Maxwell's equations in the absence of external stimuli is given by:
\begin{equation}
k^2{\bf E}-{\bf k}({\bf k}\cdot {\bf E})=\varepsilon({\bf k},\omega)\frac{\omega^2}{c^2}{\bf E}
\end{equation}

Two cases, depending on the polarization direction of the electric field vector, need to be distinguished. For transverse waves, where ${\bf k}\cdot {\bf E = 0}$, yielding the generic dispersion relation and more intriguingly for longitudinal waves, where ${\bf k}({\bf k}\cdot {\bf E})-k^2{\bf E}=0$, it requires the particular condition where $\varepsilon({\bf {k}},\omega)=0$, signifying that longitudinal collective oscillations can only occur at frequencies corresponding to zeros of the dielectric function.

Plasmon modes in doped graphene, which are obtained from the condition in which $\varepsilon({\bf k},\omega)=0$, show many special properties, and some of them are listed here.
\begin{enumerate}[leftmargin=2.3em,labelsep=4mm]
\item[(a)]
 Their momentum is larger than the light momentum with the same energy~\cite{koppens};
\item[(b)] They can be actively tuned through chemical doping or electrical gating in real time (tuning charge carriers)~\cite{Fei,Chen};
\item[(c)] They illustrate higher levels of confinement ($\lambda_{SPP}/\lambda_{light}$ $\simeq$ $0.025$ in the normal case)~\cite{koppens};
\item[(d)] They have a longer lifetime and propagating lengths ($\tau\simeq 500$ fs) \cite{lundeb};
\item[(e)] They occur in the terahertz and mid-infrared modes, which are absent in normal metals~\cite{koppens};
\item[(f)] They can be coupled with quasiparticles (for instance, generating plasmarons)~\cite{bostwick};
\item[(g)] They are used in the quasi-zero dimension as emitters~\cite{kasry, gaudreau};
\item[(h)] They show very particular properties in hybrid structures (combining graphene with other 2D crystalline materials)~\cite{principi, faridi};
\item[(k)] At long wavelength limits, they behave like $\sqrt{n^{1/2} q}$, which is proportional to the charge career as $n^{1/4}$~\cite{koppens}.
\end{enumerate}

Those properties of plasmon modes in doped graphene have been measured by using the near-field optical microscopic technique with nanometer resolution. Based on this technique, a scanning near-field optical microscope with the aperture radius much smaller than the wavelength of incident light can be used. The near-field evanescent components of light coming out from the microscope provide the required in-plane momenta. Two independent research groups performed experiments on graphene plasmonics using a similar concept~\cite{Fei,Chen} where they detected graphene surface plasmon modes. Moreover, monolayer graphene supports transverse-electric modes, which are absent in normal metals. The reason that we have this mode in doped graphene lies in the fact that the imaginary part of the conductivity is negative due to the interband transitions. Moreover, the transverse-electric modes' dispersion does not depart much from the light line, which means poor confinement, and thus, its effect is negligible.

We briefly look at another physical concept, which is the energy of the electro-magnetic field inside dispersive media, since the dielectric function is a complex-valued function. Since the amount of field localization is often quantified in terms of the electromagnetic energy distribution, a careful consideration of the effects of dispersion is necessary. In metals, the dielectric function is complex and frequency-dependent owing to the dispersion. For a field consisting of monochromatic components, Landau and Lifshitz have shown that the conservation law can be well described by an effective electric energy density $u_{\rm eff}$, defined as~\cite{loudon}:
\begin{equation}
u_{\rm eff}=\frac{\epsilon_0}{4}(\varepsilon_1+\frac{2\omega\varepsilon_2}{\gamma})|{\bf E}|^2
\end{equation}
where $\gamma=1/\tau$ in which the relaxation time of the free electron gas is $\tau$ and the dielectric function is given by $\varepsilon=\varepsilon_1+i\varepsilon_2$.

It is worth mentioning that the real part of the dielectric function or the imaginary part of the conductivity describes
the reflection of light (an elastic process); however, the imaginary part of the dielectric function describes
the absorption of light (inelastic process). Basically, plasmons are observed when the real part of the dielectric function
is negative (metallic behavior). In order to give some numbers, the plasmon modes occur at ultraviolet frequencies for aluminum and other materials,
at ultraviolet frequencies for zero- and one-dimensional carbon structures, at visible-near infrared frequencies for noble materials (Ag, Cu, Au),
at terahertz and mid-infrared frequencies for two-dimensional (2D) carbon structures (graphene) and
at amplitude modulation radio frequencies for the ionosphere.

With these general properties, let us now discuss in more detail the organization of the article. The~scope of this paper is collective modes in pristine doped two-dimensional crystalline systems, where the real part of their dielectric functions
is basically negative, with the emphasis on fundamental physics from theoretical and experimental viewpoints. Details of the band structure properties are covered in some stand stemming from the density functional theory. Phonon scattering, the effect of impurity or strain and corrugation and optical conductivity in those materials are not covered. Detailed~reviews of the plasmon modes in general are available in \cite{ Maier,Wang} and in particular for graphene in \cite{de Abajo, Jabalan, Yan, Goncalves, Tame}. Our ultimate goal is to facilitate the reader's independent study of original papers on the plasmonics of crystalline two-dimensional materials using the density functional theory.

In this article, we are using a recently-proposed
theoretical formulation based on {ab initio} density
functional theory (DFT) together with the random-phase approximation (RPA) to investigate
the electronic excitation spectrum of doped graphene, MoS$_2$ and phosphorene in monolayer and bilayer structures. To commence,
the electronic ground-state of the periodically-repeated slab of each material is first determined, and then a Dyson-like equation
is solved within the RPA to calculate the density-density
response function. Having calculated the density-density response function for each structure,
we therefore can calculate the macroscopic dielectric
function whose imaginary part gives the optical absorption spectrum,
and the collective modes are established by the zero in the real
part of the macroscopic dielectric function. The theoretical dielectric
function is related to the electron energy-loss function, and it
provides useful information about the optical properties of the
system. Here, we are just interested in the low-energy excitations
for investigating the collective modes.

\section{Theoretical Framework}\vspace{-6pt}
\subsection{Density Functional Theory}
Density functional theory (DFT) has long been the pillar of ground-state energy and density profile calculations in condensed matter science, widely used both by physicists, chemists and material researchers to study theoretically various properties of many-body systems, and in particular, it is an approach for the description of ground-state properties of metals, semiconductors and insulators. The~success of density functional theory not only encompasses standard bulk materials, but also complex materials, such as proteins and nanostructures~\cite{sholl,parr}.
\par
In 1965, Kohn and Sham proposed a practical way to implement DFT and made a significant breakthrough when they showed that the problem of many weakly-interacting electrons in an external potential can be mapped exactly to a set of non-interacting electrons in an effective external potential. The effective potential in this non-interacting particle system (the Kohn--Sham system) can be shown to be completely determined by the electron density of the interacting system and is for this reason called a density functional theory~\cite{kohn}.

\par
The Kohn--Sham (KS) equation looks like a simple one-particle Schr\"{o}dinger equation, and it can be described by the following equation:
\begin{eqnarray}
[-\frac{\hbar^2 \nabla^2}{2m}+V_{ext}(r)+V_{Hartree}+V_{xc}(r)]\Phi_{j}(r)=\epsilon_{j}\Phi_{j}(r)\nonumber
\end{eqnarray}
where $\Phi_{j}(r)$ is KS wave functions and $V_{ext}$ is the external potential acting on the interacting system. Furthermore, $V_{Hartree}$ is the Hartree part of the Coulomb electron-electron interaction.
The exchange-correlation potential, $V_{xc}(r)$, which stems from the many-body effects, describes the effects of the Pauli principle and the Coulomb potential beyond a pure classical electrostatic interaction of the electrons. Possessing the exact exchange-correlation potential means that we solve the weakly many-body problem exactly. A common approximation is the so-called local density approximation (LDA), which locally substitutes the exchange-correlation energy density of an inhomogeneous system by that of an electron gas evaluated at the local density.

\subsubsection{Computational Method}
There are several DFT packages that are available in the world, and each one mainly uses different basis sets.
PWscf, a core component of the Quantum ESPRESSO distribution~\cite{QE}, performs many different kinds of self-consistent calculations of electronic structure properties such as ground-state energy and one-electron Kohn--Sham orbitals; within density functional theory, using a plane-wave basis set and pseudopotentials.
\par
The expression of the Kohn--Sham orbitals in the plane waves basis has the form:
\begin{eqnarray}
\Phi_{n\mathbf k}(\mathbf r)=\frac{1}{\Omega}\sum_{\mathbf G}C_{n\mathbf k}(\mathbf G)e^{i(\mathbf k+\mathbf G)\cdot\mathbf r}\nonumber
\end{eqnarray}
where $\Omega$ represents the crystal volume, $G$ is the reciprocal lattice vectors and $k$ is the quasi-wavevectors of the
first Brillouin zone (BZ).
The coefficients $C_{n\mathbf k}(\mathbf G)$ are obtained by solving the LDA-KS equations self-consistently.
\par
Since the core electrons of an atom are highly localized, it would be difficult to implement them using the plane waves basis sets. Actually, a very large number of plane waves is required to expand their wave functions. Furthermore, the contribution of the core electrons to bonding compared to those of the valence electrons is usually negligible. Therefore, it is practically desirable to replace the atomic potential owing to the core electrons with a pseudopotential that has the same effect on the valence electrons~\cite{TM}.
There are mainly two kinds of pseudopotentials, norm-conserving soft pseudopotentials~\cite{vanderbilt} and Vanderbilt ultra-soft pseudopotentials~\cite{moroni}.
\par
The first part of our calculations includes determining the KS ground-state of pristine 2D crystalline materials and the corresponding wave functions and energies.
We carry out the first-principles simulations based on the DFT simulations implemented in the QUANTUM ESPRESSO code. The calculation of the density-density response functions $(\chi(q,\omega))$ is performed employing our own code. The computation of this quantity will be discussed in the next section.

\par
In this study, the electronic structures of 2D materials are computed using the Perdew--Zunger local-density approximation~\cite{pz},~unless otherwise stated.~Furthermore, we use this throughout the norm-conserving pseudopotentials and the plane wave basis. The energy convergence criteria for electronic and ionic iterations are set to be $10^{-5}$ eV and $10^{-4}$ eV, respectively.
\par
Careful testing of the effect of the cutoff energy on the total energy can be implemented to determine a suitable cutoff energy. The cutoff energy is required to obtain a particular convergence precision. Well-converged results are found with a kinetic energy cutoff equal to $50$ Ry.

\par
The $k$-point grid is another calculated parameter that must be considered, and it is used to approximate integrals of some property, e.g., the electron density over the entire unit cell. Notice~that the integration is performed in reciprocal space (in the first Brillouin zone) for convenience and efficiency. We thus use a Monkhorst--Pack~\cite{MP} $k$-point grid, which is essentially a uniformly-spaced grid in the Brillouin zone.
Geometry optimization and ground-state calculations are carried out on the irreducible part of the first BZ, using a $\Gamma$-centered and unshifted Monkhorst--Pack grid of $60\times 60\times 1$ $k$-points for graphene and MoS$_2$ and $30\times 40\times 1$ $k$-points for phosphorene. The converged electron density is then used to calculate the KS electronic structure, i.e., the single-particle energies and orbitals on a denser $k$-point mesh. We perform calculations to check the convergence of the plasmon spectra with respect to the $k$-point sampling to obtain reliable spectra, and these are listed in Table~\ref{tab:table1}.

\par
The equilibrium distance between two layers in bilayer materials is also determined by varying the interlayer distance, while keeping the in-plane lattice constant fixed at the monolayer value. In~all bilayer systems, studied in this paper, the van der Waals interaction is included in order to obtain accurate results.
\par
In the 2D case, we consider a system that is infinite and periodic only in the basal $(x-y)$ plane, but confined along the third $(z)$ direction. A so-called supercell approach is commonly used to treat 2D systems, in which the system is modeled by repeated 2D slabs, separated by a large vacuum region along the $z$ direction.
We use a vacuum region of at least {$20$ \AA}~to avoid spurious interaction between the periodic images. We have discovered that increasing these separations would not affect the band structure of the system. The structural parameters, band gap and sampling of the reciprocal space BZ to calculate the density-density response function are listed in Table~\ref{tab:table1}.

\begin{table}[H]
\caption{\label{tab:table1} The lattice constants ($a, b$) and band gap (in eV) of different 2D crystalline materials. Sampling of the reciprocal space Brillouin zone (BZ) is done by a Monkhorst--Pack (MP) grid for the considered 2D materials.}

\centering
\begin{tabular}{cccccccc}
\toprule
\multirow{2}{*}{\textbf{{2D Structures}}\vspace{-3pt}} & \multirow{2}{*}{\textbf{{Lattice}}\vspace{-3pt}} &\multirow{2}{*}{ \textbf{{a (\AA)}} \vspace{-3pt} }& \multirow{2}{*}{\textbf{{b (\AA)}}\vspace{-3pt}} &\multicolumn{2}{c}{\textbf{Gap (eV)}}& \multicolumn{2}{c}{\textbf{MP}}\\ \cmidrule{5-8}
& & & & \textbf{Monolayer} & \textbf{Bilayer} & \textbf{Monolayer} & \textbf{Bilayer}\\
\midrule
 graphene  & Hexagonal & $2.46$   & $-$  & $0.00$& $0.00$ & $101\times101\times1$ & $201\times201\times1$  \\
 MoS$_2$  & Hexagonal & $3.15$   & $-$  & $1.80$& $1.05$ & $101\times101\times1$ & $151\times151\times1$  \\
 phosphorene & Rectangular & $4.62$   & $3.30$ & $0.98$& $0.63$ & $~60\times~~80\times1$ & $120\times~160\times1$ \\
\bottomrule
\end{tabular}
\end{table}

\subsection{Density-Density Response Function}
\par
A central quantity in the theoretical formulation of the many-body
effects in electronic systems is the non-interacting dynamical polarizability function~\cite{Vignale}
$\chi^{(0)}({\bf q},\omega)$ for a finite chemical
potential, $\mu$. Here, we would like to emphasize that we have calculated the non-interacting polarization function for doped graphene, MoS$_2$ and phosphorene, both monolayer and bilayer structures based on DFT simulations.

The expression of the non-interacting density-density response function of a three-dimensional periodic electrons in the reciprocal space is:
\begin{eqnarray}
\chi_{\mathbf G \mathbf G'}^{0}(\mathbf q,\omega)=\frac{2}{\Omega}\sum_{\mathbf k,\nu,\nu'}\frac{f_\nu(\mathbf k)-f_{\nu'}(\mathbf k+\mathbf q) }{\hbar\omega+\varepsilon_n(\mathbf k)-\varepsilon_m(\mathbf k+\mathbf q)+i\eta}
\rho_{\nu\nu'}^{\mathbf k\mathbf q}(\mathbf G) \rho_{\nu\nu'}^{\mathbf k\mathbf q}(\mathbf G')^*
\label{response}
\end{eqnarray}
which is obtained from the Adler--Wiser periodic system~\cite{kubo}.
Here, $\varepsilon_n(\mathbf k)$ and $\varepsilon_m(\mathbf k+\mathbf q)$ denote the empty and filled bands and $f_n(\mathbf k)=\theta(E_{\rm F}-\varepsilon_n(\mathbf k))$ is the Fermi-Dirac distribution of the charge carrier with energy $\varepsilon_n(k)$ at temperature zero. Furthermore, $G$ and $G'$ are the three-dimensional (3D) reciprocal lattice vectors, and $\omega$ is the frequency.

In this theory, the linear combination of plane-waves is used to determine the KS single-particle orbitals of the DFT. The KS wave functions are normalized to unity in the crystal volume $\Omega$. The sum is over a special set of $k$ vectors and energy bands ($\nu$ and $\nu'$). In Table~\ref{tab:table1}, the sampling $k$ point in the BZ is indicated for different 2D materials in order to fully converge the results.
The factor of two accounts for the spin degeneracy, and $\eta$ is a small (positive) lifetime broadening parameter.

The matrix elements of Equation (\ref{response}) have the form:
\begin{eqnarray}
 \rho_{\nu\nu'}^{\mathbf k\mathbf q}(\mathbf G)=<\Phi_{\nu\mathbf k}|e^{-i(\mathbf q+\mathbf G)\cdot \mathbf r}|\Phi_{\nu'\mathbf k+\mathbf q}>_{\Omega}
\end{eqnarray}
where $\mathbf q$ is the momentum transfer vector parallel to the $x-y$ plane.
Wave functions $\Phi_{n\mathbf k}(\mathbf r)$ are the KS electron wave functions and when
expanded in the plane-wave basis have the form:
\begin{eqnarray}
\Phi_{n\mathbf k}(\mathbf r)=\frac{1}{\Omega}\sum_{\mathbf G}C_{n\mathbf k}(\mathbf G)e^{i(\mathbf k+\mathbf G)\cdot\mathbf r}
\end{eqnarray}
where the coefficients $C_{n\mathbf k}(\mathbf G)$ are obtained by solving the LDA-KS equations self-consistently.

\par
The exact interacting density-response function can be obtained in the framework of the DFT, as~follows~\cite{DFT}:
\begin{eqnarray}
\chi_{\mathbf G\mathbf G'}=\chi^0_{\mathbf G\mathbf G'}+\sum_{{\bf G_1 G_2}}\chi^0_{\mathbf G\mathbf G_1}\nu_ {\mathbf G_1\mathbf G_2}\chi_{\mathbf G_2\mathbf G'}
\end{eqnarray}
where $\nu_{\mathbf G\mathbf G'}$ represents the Fourier coefficients of an effective electron-electron interaction. In the electron liquid, the bare Coulomb interaction is given by $\nu_{\mathbf G\mathbf G'}^0=4\pi e^2\delta_{\mathbf G\mathbf G'}/\epsilon|\mathbf q+\mathbf G|^2$ where $\epsilon$ is the average dielectric constant of the environment.
In all our numerical results, we consider $\epsilon=1$.
The RPA procedure, an approximation valid in the high-density limit, takes into account electron interaction only to the extent required to produce the screening field, and thus, the response to
the screened field is measured by $\chi^0$. The RPA follows from a microscopic approach whose main assumption is that the electrons respond not to the bare Coulomb potential, but to an effective potential resulting from the dynamical rearrangement of charges in response to the Coulomb potential.
The long-range behavior of the Coulomb interaction allows non-negligible interactions between repeated planar arrays even at a large distance. This unphysical phenomenon can be removed by replacing $\nu_{\mathbf G\mathbf G'}$ with the truncated Fourier
integral over the cutoff plane axis $(z)$ \cite{pisarra,vacac}, and thus, we have:
\begin{eqnarray}
\nu^0_{\mathbf G\mathbf G'}=\frac{2\pi e^2\delta_{\mathbf g\mathbf g'}}{|\mathbf q+\mathbf g|}\int_{-L/2}^{L/2}dz\int_{-L/2}^{L/2}dz'e^{i(G_zz-G_{z}'{z}')-|\mathbf q+\mathbf g||z+z'|}
\end{eqnarray}
where the $\mathbf g$ and $G_z$ denote the in-plane and out-plane components of $\mathbf G$, and we assume that $q$ is never zero owing to a uniform background of positive charge.

In the framework of the linear response theory, the inelastic cross-section corresponding to a process where the external perturbation creates an excitation of energy $\hbar \omega$ and wavevector $\mathbf q+\mathbf G$ is related to the diagonal elements of the dielectric function in the level of the RPA:
\begin{eqnarray}
\varepsilon_{\mathbf G\mathbf G'}=\delta_{\mathbf G \mathbf G'}-\sum_{\mathbf G_1} \nu^0_{\mathbf G\mathbf G_1}\chi^0_{\mathbf G_1 \mathbf G'}
\end{eqnarray}
and the plasmon modes are established by the zero in the real part of the macroscopic dielectric function given by:
\begin{eqnarray}\label{epsilon}
\varepsilon(q,\omega)=\frac{1}{(\varepsilon^{-1})_{\mathbf G\mathbf G'}}|_{\mathbf G=\mathbf G'=0}\label{eq11}
\end{eqnarray}
as long as there is no damping.
The electron-energy loss (EEL) is proportional to the imaginary part of the inverse dielectric function, which is given by:
\begin{eqnarray}\label{loss}
E_{EEL}(q,\omega)=-\Im m[1/\varepsilon(q,\omega)]
\end{eqnarray}

It is worth mentioning that the nonlocal field effects are included in EEL through the off-diagonal elements of the general $\chi_{\mathbf G \mathbf G'}$~\cite{nlc} function. In addition, one may use the non-local dynamical conductivity to describe the electronic processes and light-matter interactions. This can be simply achieved by writing the longitudinal conductivity in terms of the density-density response function through the relation:
\begin{eqnarray}\label{conductivity}
\sigma(q,\omega)=\frac{i e^2 \omega}{q^2}\chi(q,\omega)
\end{eqnarray}

We note that Equations (\ref{epsilon})--(\ref{conductivity}) are the fundamental physics describing the optoelectronic interactions in 2D crystalline materials.

\section{Result and Discussion}
\vspace{-6pt}
\subsection{Monolayer Graphene}
\begin{figure}[H]
\centering
\includegraphics[width=9 cm]{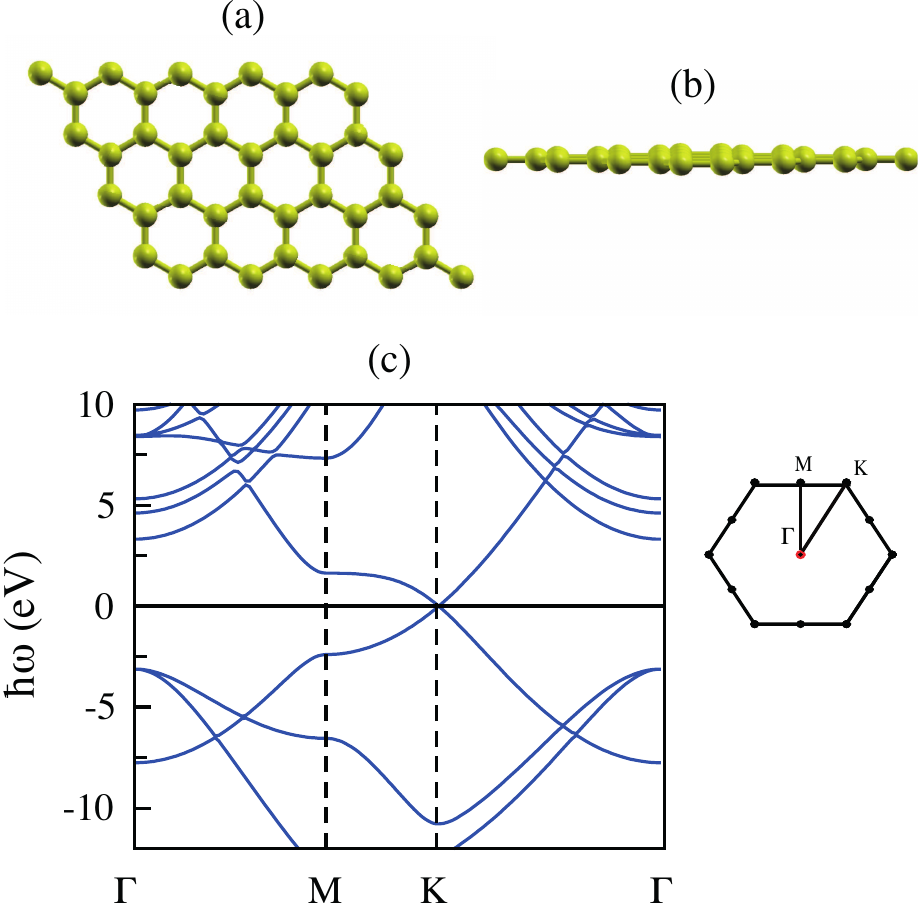}
\caption{(\textbf{a},\textbf{b}) Top and side view of monolayer graphene. (\textbf{c}) The band structure of graphene along the high symmetry $\Gamma- M- K- \Gamma$ directions and the associated Brillouin zone. The zero in the energy axis is set at the Fermi level as shown by the solid line.}\label{fig1}
\end{figure}

Graphene is a 2D layer of carbon atoms arranged in a honeycomb lattice with Dirac cones, i.e., massless Dirac fermion at $K$ point, where the $\pi$ and $\pi ^*$ bands show a linear energy dispersion \cite{rostami,polini,neto}. In Figure~\ref{fig1}, we illustrate the atomic structure and also the electronic energy band structure of graphene based on our DFT simulations.

\par
Research on collective electronic excitations (plasmons) in graphene has attracted enormous interest both from theoretical and experimental viewpoints~\cite{despoja2012,yan,wunsch,jablan,wach,koppens,gao,lundeb,novko,stauber}.
Three kinds of the collective excitations of electrons can be considered in graphene that unroll on a wide range of energy. The first kind is attributed to finite electron doping, originating from the intraband transitions of Dirac fermions in the vicinity of the $K$ and $K'$
 points of the BZ at low energies (0--2 eV), and it can be regarded as intraband plasmon modes. The second kind of plasmons in the monolayer graphene is the intrinsic $\pi$ plasmons, arising from the collective excitations of electrons from the $\pi$ to $\pi^*$ bands at energies of about 4--15~eV. At~higher energies, $\sigma$ bands start to contribute, and the mixture of the $\pi\rightarrow \pi^*$ and $\sigma$ transitions leads to another kind of plasmon excitation of graphene, usually denoted as $\pi+\sigma$ plasmons. It is worthwhile mentioning here that the calculation of the $\pi+\sigma$ plasmon would require including high-lying bands; hence, the {ab initio} approach would be the more appropriate way to capture them~\cite{li}.

The plasmon spectrum of a pristine single-layer graphene was investigated in \cite{eberlein} using {ab~initio} calculations. They observed the $\pi$ and $\pi+\sigma$ plasmon modes in freestanding single sheets at $4.7$~and~$14.6$ eV, which were red shifted in comparison to the corresponding modes in the bulk graphite.
\par
Our attention is now focused on the intraband and $\pi$ plasmons of the spectrum (below $10$ eV), and it appears to be owed to the intraband and interband transitions, respectively, as mentioned above.
\par
\begin{figure}[H]
\centering
\includegraphics[width=7.5 cm]{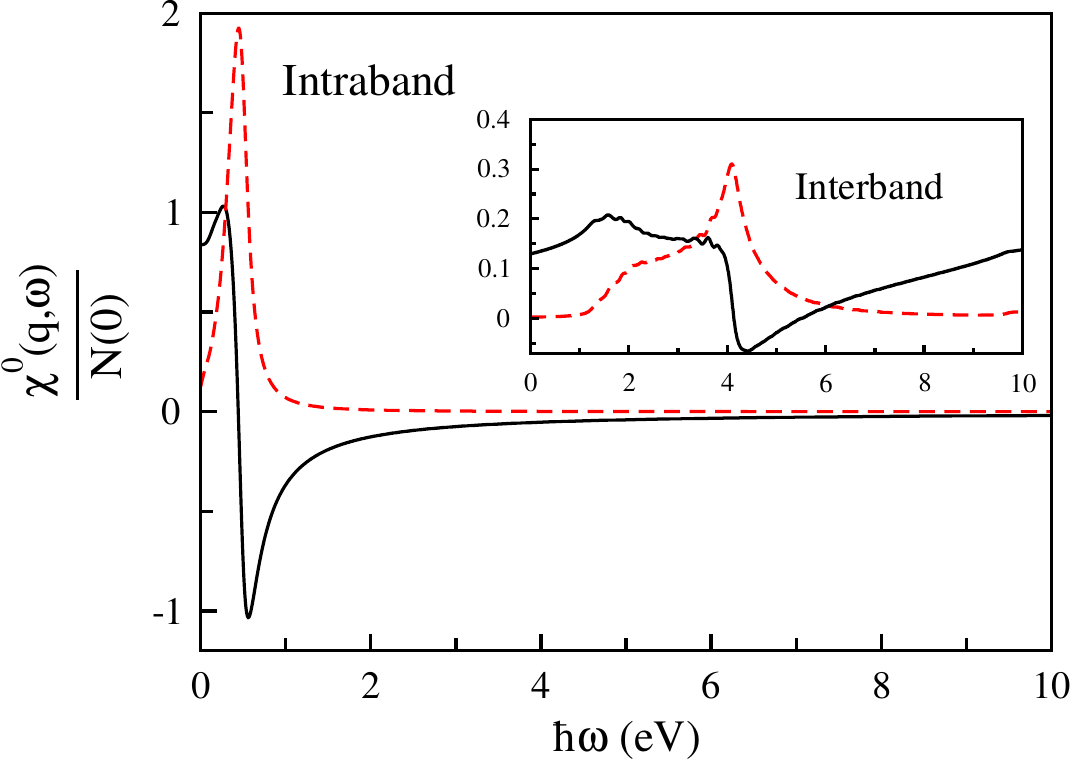}
\caption{The real (solid line) and imaginary (dashed line) parts of the non-interacting density response function of monolayer graphene in units of the Fermi-level density of states as a function of $\hbar \omega$ for $q=0.076$ \AA$^{-1}$. }\label{fig2}
\end{figure}

In order to better perceive the electronic excitations of monolayer graphene, we calculate the non-interacting density-density response function, $\chi^0 (q,\omega)$.
The non-interacting density-density response function can be decomposed into two parts where the first part can be considered by only including transitions within a band and the second part contains all transitions between separate bands.
An example of the non-interacting density-density response function is shown in Figure~\ref{fig2}, for~$q=0.076$ \AA$^{-1}$ and $E_{\rm F}= 0.8$ eV (the Fermi level is shifted upward by $0.8$ eV above the Dirac cone, and~this corresponds to an electron doping level of $7.4\times 10^{13}\,$cm$^{-2}$). The $\lim_{q \to 0} \chi^0(\mathbf q, \omega=0$) is~finite and equal to the density of states at the Fermi energy, $N(0)$ a measure of the number of excited states. This figure shows that intraband transitions contribute more at low-energy, while the interband transitions dominate at high-energy.

\begin{figure}[H]
\centering
\includegraphics[width=6.7 cm]{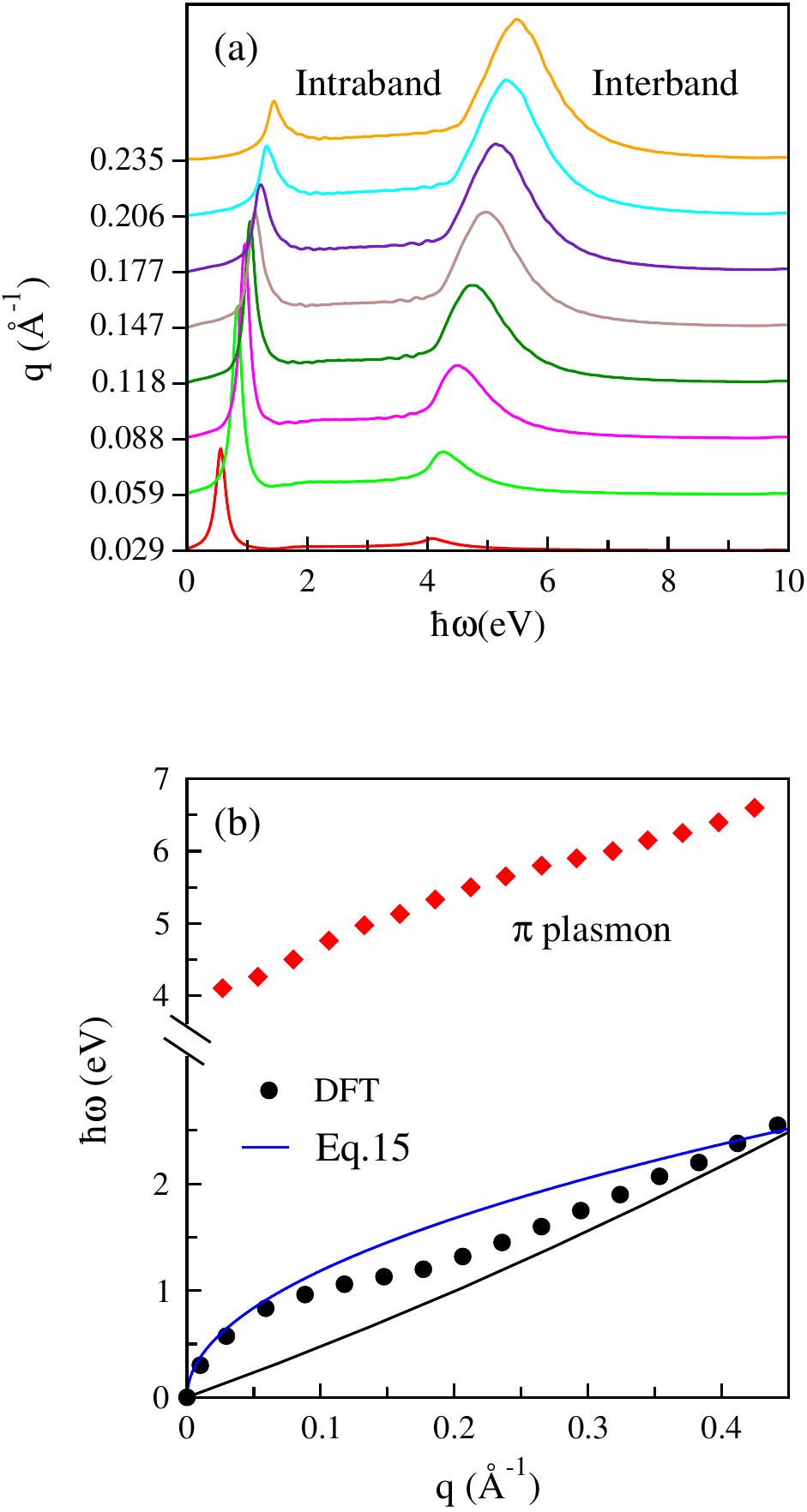}
\caption{(\textbf{a}) The electron-energy loss (EEL) function for different momentum transfers and (\textbf{b}) plasmon dispersions of electron-doped graphene along the $\Gamma-M$ direction. In this case, the Fermi level is shifted upward by $0.8$ eV above the Dirac cone, and this corresponds to a doping level of $7.4 \times 10^{13}$\,cm$^{-2}$. The plasmon mode based on the semiclassical approach given by Equation (\ref{eq15}) is plotted. The boundary of the electron-hole continuum is indicated with a black line.}\label{fig3}
\end{figure}

In Figure~\ref{fig3}, the EEL functions for various momentum transfers and the plasmon spectrum of electron doped monolayer graphene ($n=7.4\times 10^{13}\,$cm$^{-2}$) are computed using the linear response DFT-RPA approach. The effect of doping on the band structures of 2D materials is ignorable~\cite{despoja}. In~the long-wavelength limit, plasmons can be viewed as a center-of-mass oscillation of the
electron gas as a~whole. The physical origin of plasmons is described as follows. When electrons in free space move to screen
a charge inhomogeneity, they tend to overshoot the mark. They are then pulled back toward the charge disturbance
and overshoot again, setting up a weakly-damped oscillation. The restoring force responsible for the oscillation is the
average self-consistent field created by all the electrons. As~expected, the dispersion behavior of the plasmon mode at the low-energies shows a standard $\sqrt q$ dispersion predicted in the 2D electron gas system. To be more precise, the plasmon mode at the long wavelength limit in graphene is given by:
\begin{equation}
\omega_p(q)=\sqrt{\frac{2D_0}{\epsilon}q}\left[1+\frac{12-N^2_f\alpha_{ee}^2-8(1-\kappa_0/\kappa)}{16}\frac{q}{q_{TF}}+...\right]
\end{equation}
where $D_0= v_{\rm F}k_{\rm F} e^2/\hbar$ is the Drude weight in graphene. For ordinary parabolic-band fermions with mass $m_b$, the Drude weight is given by $D_0=\pi n e^2/m_b$. Graphene's fine-structure constant is $\alpha_{ee}=e^2/\hbar v_{\rm F} \epsilon$; $N_{F}=4$ is the flavor number in graphene; the Thomas--Fermi screening wave number is defined by $q_{TF}=N_F \alpha_{ee} k_{\rm F}$; and
the electron compressibility of interacting and non-interacting graphene is $\kappa$ and $\kappa_0$, respectively~\cite{lundeb}
. The second term in the square brackets refers to the quantum non-local effects. Notice that in the classical picture, the long-wavelength plasmon mode behaves like $\omega_p(q)=\sqrt{\frac{2\pi n e^2}{\epsilon m}q}$, which is totally different from the one we have in graphene.

Our numerical results show that the general behavior of the plasmonic dispersion agrees with the results from \cite{despoja,li}. The dispersion relation of the $\pi$ plasmon is presented, and it has a quasi-linear behavior, in good agreement with the earlier theoretical study~\cite{li}. In this figure, the electron-hole continuum is indicated with a black line, and it can be obtained at any specific momentum transfer $\mathbf q$ by the difference between system energy at $\mathbf{k_{\rm F}}$ and $\mathbf{k_{\rm F}+q}$.

We would like to compare the plasmon mode obtained here with that calculated within a~semiclassical approach. We suppose that the graphene sheet, along the $x$ and $y$ directions, is located between two semi-infinite dielectric media with the same dielectric constant, $\epsilon$, and consider a solution of Maxwell's equations for a transverse magnetic wave. By using the proper boundary conditions, we~arrive at~\cite{Goncalves}:
\begin{equation}
2\frac{\epsilon}{\sqrt{q^2-\omega^2\epsilon/c^2}}=-i\frac{\sigma(\omega)}{\omega\epsilon_0}\label{eq15}
\end{equation}
which describes the plasmon mode, $\omega(q)$, of graphene with conductivity $\sigma(\omega)$. It should be noted that this expression implements every 2D crystalline material with its $\sigma(\omega)$. This~equation has solutions, if the imaginary part of the conductivity is positive and its real part vanishes. The~conductivity of graphene from the terahertz to mid-infrared regime is dominated by a Drude term and given by:
\begin{equation}
\sigma(\omega)=\frac{e^2}{\pi \hbar}\frac{iE_{\rm F}}{\hbar \omega}
\end{equation}

It turns out that at the long wavelength limit, the plasmon mode is given by:
\begin{equation}
\hbar \omega_p=\sqrt{\frac{2e^2}{\epsilon_0}E_{\rm F} q}
\end{equation}
which is the same expression that we have in the quantum many-body framework. We also include the semiclassical plasmon mode in Figure~\ref{fig3}b, and our numerical results show that they are the same at long wavelength regimes; however, at large momenta, the semiclassical plasmon mode deviates from that calculated by the many-body approach, especially at lower electron density.

\subsection{Bilayer Graphene}

\begin{figure}[H]
	\centering
	\includegraphics[width=12 cm]{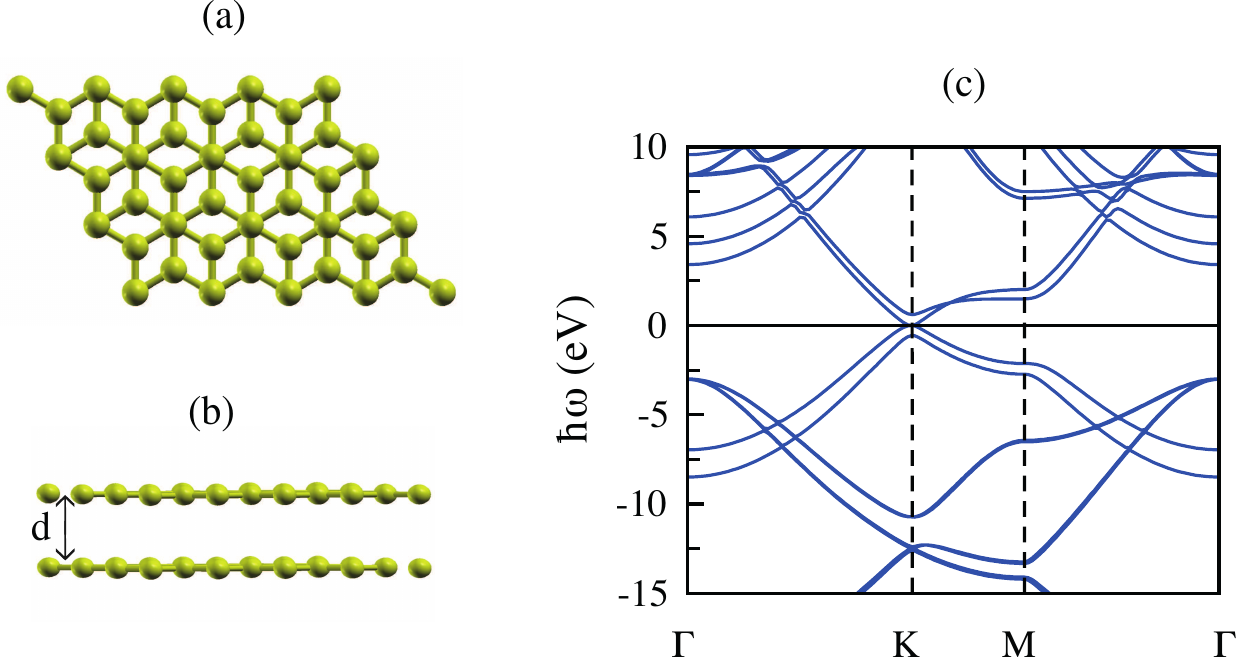}
	\caption{(\textbf{a},\textbf{b}) Top and side view of the atomic structure of bilayer graphene (BLG). (\textbf{c}) The band structure of BLG along the high symmetry $\Gamma- K- M- \Gamma$ directions in reciprocal space. The optimized interplane distance ($d$) is calculated to be $3.45$ \AA.}\label{fig4}
\end{figure}

Bilayer graphene displaces the simplest possible system where graphene sheets are brought together to create a new nanostructure whose physical properties show remarkable similarities and differences as compared to monolayer graphene.
Bilayer graphene, like single-layer graphene, is a zero-gap semimetal that consists of two coupled monolayers of carbon atoms stacked as in natural graphite (AB stacking or Bernal-stacked form where half of the atoms lie directly over the center of a hexagon in the lower graphene sheet, and half of the atoms lie over an atom) yielding a unit cell of four atoms (Figure~\ref{fig4})~\cite{mcCann,min,ohta}.

We calculate the band structure of bilayer graphene through DFT simulations, and this is illustrated in Figure~\ref{fig4}.
Bilayer graphene has four electronic bands (a pair of conduction bands and a pair of valence bands) with $p_z$ symmetry, namely: $\pi_1$, $\pi_2$, $\pi_1^*$ and $\pi_2^*$. The dispersion of these bands is parabolic near the $K$ point, and their occupation depends on the doping values. In undoped bilayer graphene, the $\pi_1$ and $\pi_2$ bands are fully occupied, while the $\pi_1^*$ and $\pi_2^*$ bands are fully unoccupied. Furthermore, an overlap of the $\pi_1^*$ and $\pi_2^*$ bands along $K-M$ leads to an anisotropic dispersion at energies from $1$--$1.5$ eV.

Research on the collective electronic excitations (plasmons) in freestanding multilayer graphene~\cite{wach2014,borghi,profumo} shows that
, similar to single-layer graphene and graphite, the spectra of bilayer graphene feature two characteristic high-energy plasmon peaks, i.e., the $\pi$ plasmon below $10$ eV and the $\pi+\sigma$ plasmon above $15$ eV, but these plasmons are red shifted with respect to graphite. In the~low-energy range, a conventional 2D plasmon was predicted to exist originating from the intraband~transitions.
\par
It is noteworthy to mention that, as reported in~\cite{sarma}, when the two layers are near each other (separated by a distance $d$ in the $z$ direction with the 2D layers in the $x-y$ 
 plane), the 2D plasmons are coupled by the interlayer Coulomb interaction leading to the formation of in-phase and out-of-phase interlayer density fluctuation modes: an in-phase optical plasmon mode, where the densities in the two layers fluctuate in phase with the usual 2D plasma dispersion ($\omega\sim\sqrt q$) and an out-of-phase acoustic plasmon mode, where the densities in the two layers fluctuate out-of-phase with a linear wavelength dispersion ($\omega\sim q$).
\par
In a recent study, the plasmon modes of undoped (intrinsic) and doped (extrinsic) bilayer graphene were calculated and analyzed carefully based on density-functional theory in an energy range from a~few meV to $\sim$30 eV, along the inequivalent $\Gamma-K$ and $\Gamma-M$ directions~\cite{pisarra}.
In that paper, they found an acoustic plasmon mode for momenta along the $\Gamma-M$ direction for a positive shift in the Fermi level of $1$ eV. Although they have claimed this acoustic plasmon is an undamped collective excitation, an overdamped acoustic plasmon was predicted by Das Sarma {et al.}~\cite{sarma}.
\par

\begin{figure}[H]
	\centering
	\includegraphics[width=7 cm]{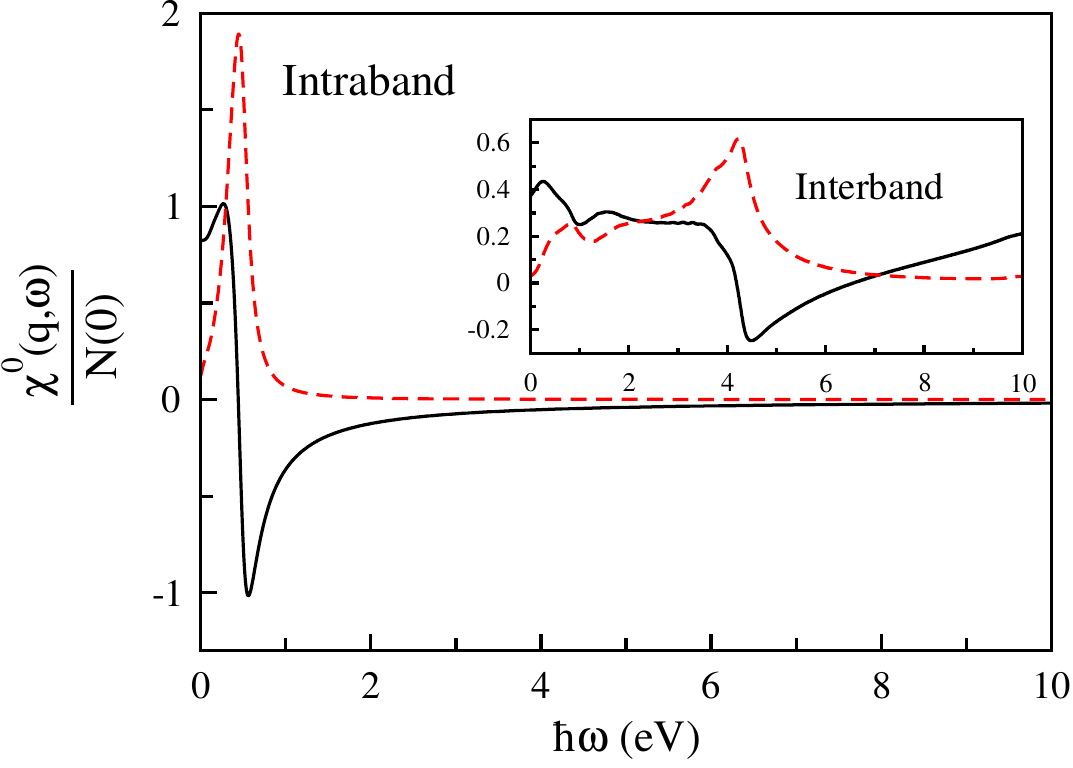}
	\caption{The inter- and intra-band terms of the non-interacting response function ($\chi^0 (q,\omega)$) of doped bilayer graphene in units of the Fermi-level density of states as a function of $\hbar \omega$ for $q=0.084$ \AA$^{-1}$ along the $\Gamma-M$ direction. The real and imaginary parts of this function are indicated by solid and dashed lines,~respectively.}\label{fig5}
\end{figure}

To investigate the origin of different plasmon modes in bilayer graphene, we calculate the interband and intraband parts of the non-interacting response function of bilayer graphene, and we present our results in Figure~\ref{fig5} for only $q=0.084$ \AA$^{-1}$. In this case, the Fermi energy is shifted upward by $0.7$ eV towards the bottom of the conduction bands in bilayer graphene, corresponding to $n=9.5\times 10^{13}$ cm$^{-2}$. Our numerical results show that the non-interacting response function of bilayer graphene is similar to its monolayer, but there is an extra contribution of the interband part at low-energies that is obviously absent in monolayer graphene.

\begin{figure}[H]
	\centering
	\includegraphics[width=6.5 cm]{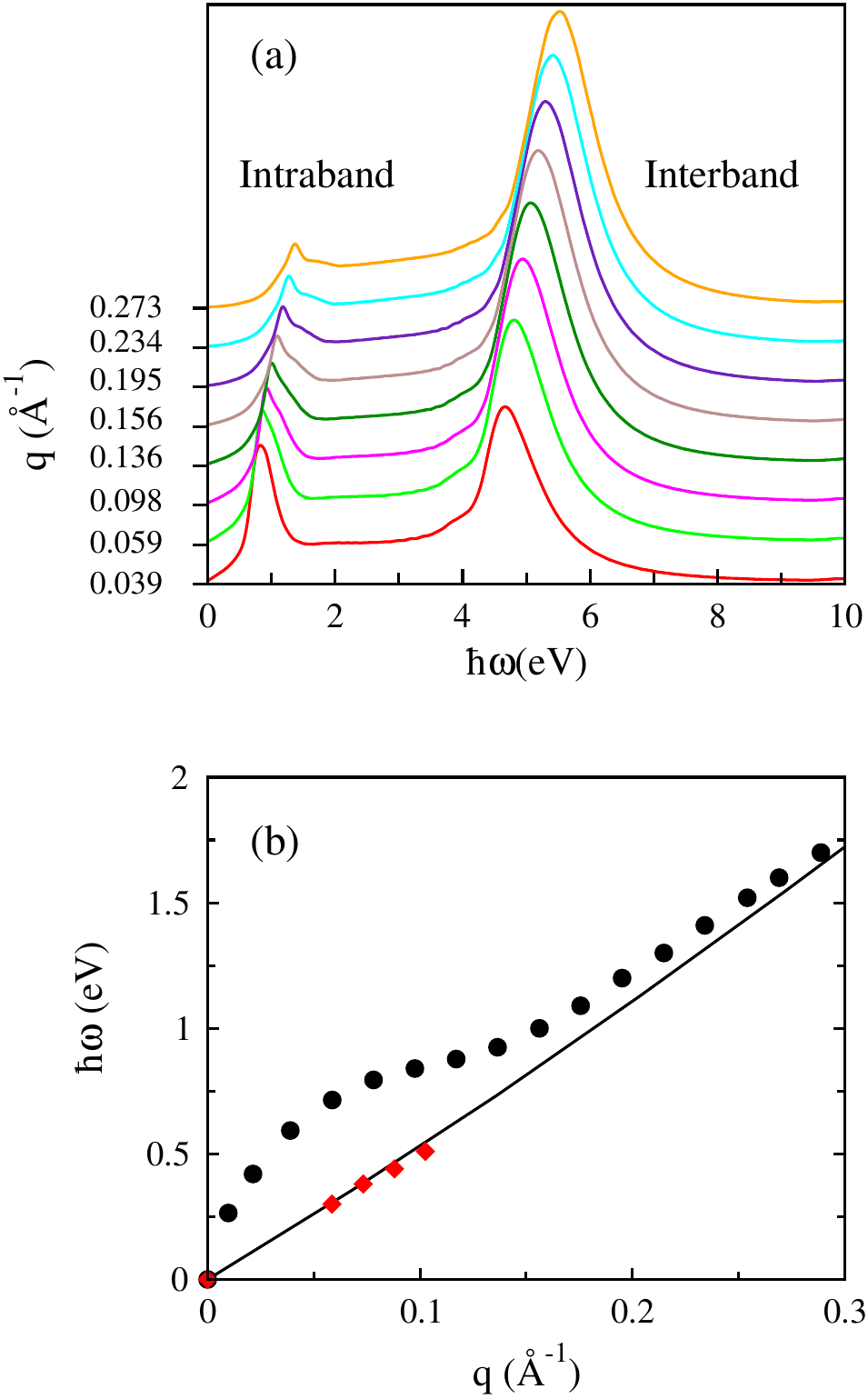}
	\caption{(\textbf{a}) The loss spectra for different values of $q$ and (\textbf{b}) the plasmon dispersion of doped bilayer graphene with $E_{\rm F}= 0.7$ eV corresponding to $n=9.5\times10^{13}$\,cm$^{-2}$ along the $\Gamma-M$ direction. The~diamond symbols refer to the acoustic plasmon mode.}\label{fig6}
\end{figure}

\textls[-18]{In Figure~\ref{fig6}, we illustrate the loss spectra for various amounts of $q$ and also electronic excitations for bilayer graphene along the $\Gamma-M$ direction for $n=9.5\times 10^{13}$ cm$^{-2}$. The loss spectra involves both intraband and interband excitations at energies below $10$ eV, but we neglect to show $\pi$ plasmon in~Figure~\ref{fig6}b.}

It is clear that the plasmon dispersion of bilayer graphene follows a general $\sqrt q$ dispersion at low energies, which is also seen in monolayer graphene. These results are in reasonable agreement with earlier work reported by Pisarra and coauthors~\cite{pisarra}. Our {ab initio} calculations show the acoustic plasmon mode, which is highly damped in bilayer graphene structures.

Most importantly, by breaking the inversion symmetric of the two layers, a non-zero band gap can be induced. The potential of a
continuously tunable band gap through a gate voltage applied perpendicularly to the sample is very
interesting~\cite{mccann_PRB_06, min, guinea_prb_06}. The realization of a widely tunable electronic band gap in electrically-gated bilayer graphene has been experimentally demonstrated~\cite{zhang_nature_09}. They~showed a gate-controlled, continuously tunable band gap of up to
$250$ meV by using dual-gate bilayer
graphene field-effect transistor infrared micro-spectroscopy. Moreover, this electrostatic band gap control suggests nanoelectronic and nanophotonic device applications based on
graphene. Notice~that the band gap can be observed in photoemission, magneto transport, infrared spectroscopy and scanning tunneling
spectroscopy.

The low-energy effective model Hamiltonian for a gated bilayer graphene
can be written as:
\begin{equation}\label{biased}
H_{\rm eff}=-\frac{1}{2m} (\vec{\sigma}\cdot \vec{p})\sigma_x (\vec{\sigma}\cdot \vec{p})+\Delta\sigma_z
\end{equation}
where ${\vec \sigma}$ are Pauli matrices and $\Delta$ is the gated energy. In bilayer graphene, changing the applied gate voltages will turn into controlling the electron density, $n$, and the interlayer asymmetry different potential energies, $\Delta$. In other words, the asymptotic energy $\Delta V$ is related to layer density, and the layer densities depend on $\Delta$,
ultimately~\cite{asgari-book}. Here, we ignore this effect and assume that the band gap is independent of the electron density.

\begin{figure}[H]
	\centering
	\includegraphics[width=7.2 cm]{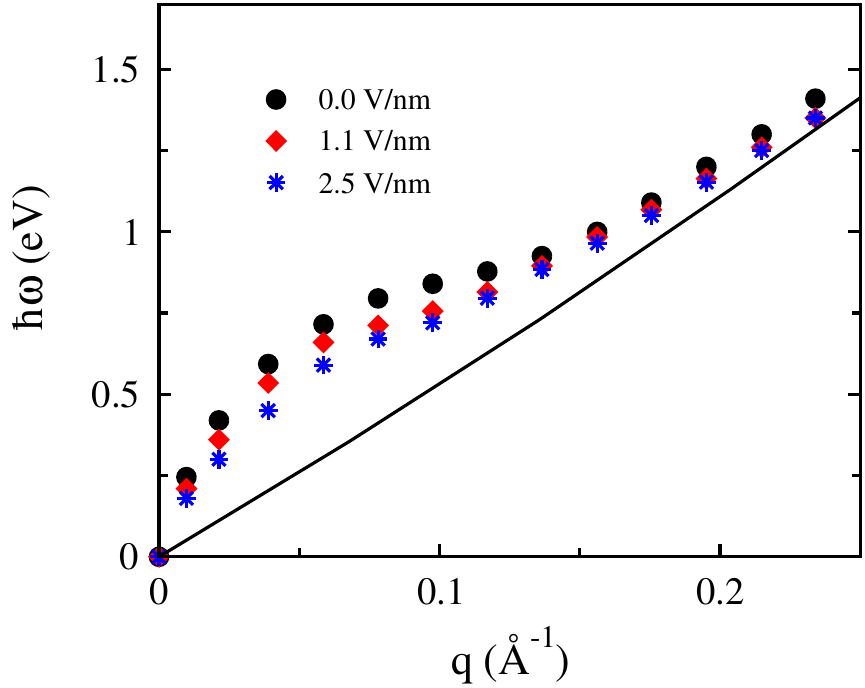}
	\caption{The gate voltage dependence of the plasmon mode in bilayer graphene. The values of the gate voltages are $1.1$ V/nm and $2.5$ V/nm related to band gaps of about $0.1$ and $0.2$ eV, respectively.}\label{fig71}
\end{figure}

We have examined first-principle calculations to investigate the plasmon modes of AB stacked bilayer graphene in the presence of a perpendicular applied electric field. Figure~\ref{fig71} shows the gate dependence of the optical plasmon modes of bilayer graphene for $\Delta=1.1$ and 2.5 eV. Clearly, the gap reduces the plasmon mode and softens the collective excitation modes, especially at the long wavelength limit. In this regime, the plasmon mode behaves slightly different from that obtained for a~system with $\Delta=0$ given by $\omega_p(q)=\sqrt{\frac{e^2}{4\epsilon_0}\frac{n q}{m}}$~\cite{Chakraborty}. Moreover, the interband contribution
decreases with the gated energy $\Delta$; however, the intraband contribution increases with
the bias owing to the fact that the energy dispersion leads to an enhancement of the density of states near the Fermi energy.

\subsection{Monolayer MoS$_2$}
The lack of a natural band gap makes graphene unsuitable for developing optoelectronic and photovoltaic devices~\cite{bonacco}. A particularly interesting class of 2D materials is the transition metal dichalcogenides (TMDC) whose electronic properties range from semiconducting to metallic and even superconducting. Among them, molybdenum disulfide (MoS$_2$) with a natural band gap is gaining increasing interest~\cite{rukelj}.
MoS$_2$ has a hexagonal structure that consists of two planes of hexagonally-arranged S atoms bonded through covalent bonds to central layer Mo atoms.

\par
\begin{figure}[H]
\centering
\includegraphics[width=9 cm]{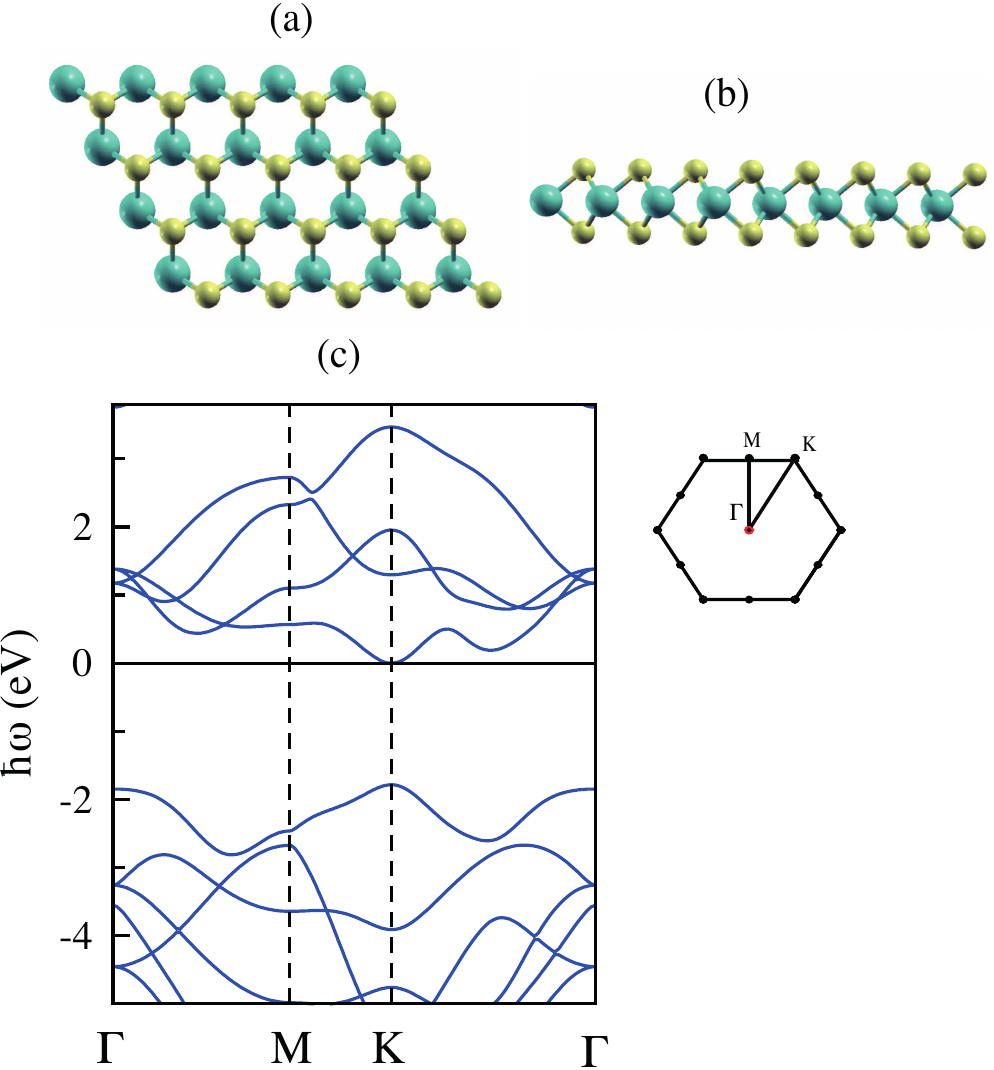}
\caption{\textls[-5]{(\textbf{a},\textbf{b}) Top and side view of the atomic structure of molybdenum disulfide (MoS$_2$). (\textbf{c})~The~band structure of MoS$_2$ along the high symmetry $\Gamma- M- K- \Gamma$ directions and the associated Brillouin zone. The Fermi level is set at 0 eV. The blue balls are molybdenum atoms, and the yellow ones are sulfur~atoms.}}\label{fig7}
\end{figure}

In agreement with previous reports~\cite{kadantsev,kumara,rostami2,rostami3}, the MoS$_2$ monolayer is a direct band gap semiconductor with a maximum of the valence and the minimum of the conduction band located at the K
 point of the Brillouin zone (Figure~\ref{fig7}). The bands around the energy gap are relatively flat, which are as expected from the $d$ character of the electron states at these energies. The states around the Fermi energy are mainly due to the $d$ orbitals of Mo, while strong hybridization between $d$ orbitals of Mo and $p$ orbitals of $S$ atoms below the Fermi energy has been observed~\cite{Vignale}.

The GW approximation, is an approximation made in order to calculate the self-energy of a many-body system of electrons, has been shown to provide very reliable descriptions of the electronic and dielectric properties for many semiconductors and insulators~\cite{shishkin}.
The recent quasiparticle self-consistent GW calculations have reported that MoS$_2$ is a direct gap semiconductor at both the LDA and GW levels, and the GW gap is $2.78$ eV at both the $K$ and $K'$
 points~\cite{Qiu}. In the following, we will explore the dispersion behavior of plasmon modes of monolayer MoS$_2$ by using our DFT-RPA code.

\par
The intraband plasmons in metallic single-layer transition metal dichalcogenides (TMDCs) have been studied using density functional theory in the random phase approximation~\cite{andersen}. They have found that at very small momentum transfer, the plasmon energy follows the classical $\sqrt q$ behavior of free electrons in 2D. For larger momentum transfer, the plasmon energy is significantly red shifted due to screening by interband transitions.
\par
\begin{figure}[H]
\centering
\includegraphics[width=7 cm]{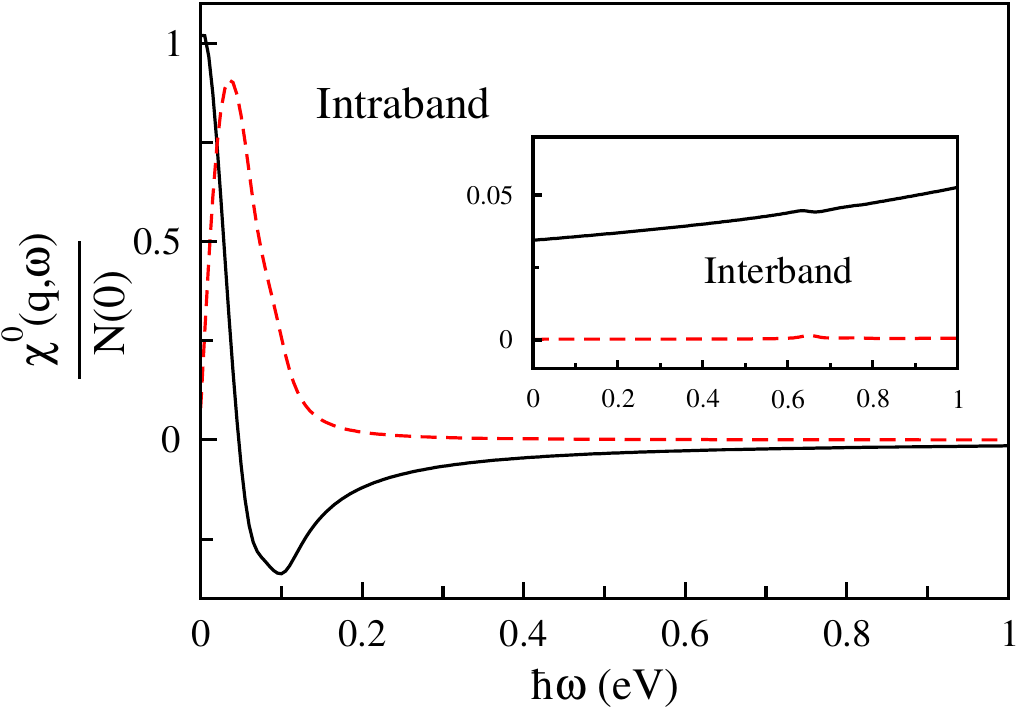}
\caption{The interband and intraband of non-interacting response function ($\chi^0 (q,\omega)$) of doped MoS$_2$ in units of the Fermi-level density of states as a function of $\hbar \omega$ for $q=0.069$ \AA$^{-1}$.}\label{fig8}
\end{figure}

The non-interacting response function of MoS$_2$ with $n=5.6\times10^{13}\,$cm$^{-2}$ and for $q=0.069$ \AA$^{-1}$ is plotted in Figure~\ref{fig8} and shows that the intraband term of $\chi^0 (q,\omega)$ is dominated at energies below $1.0$~eV, and the interband term is inconsiderable.

\begin{figure}[H]
\centering
\includegraphics[width=6.7 cm]{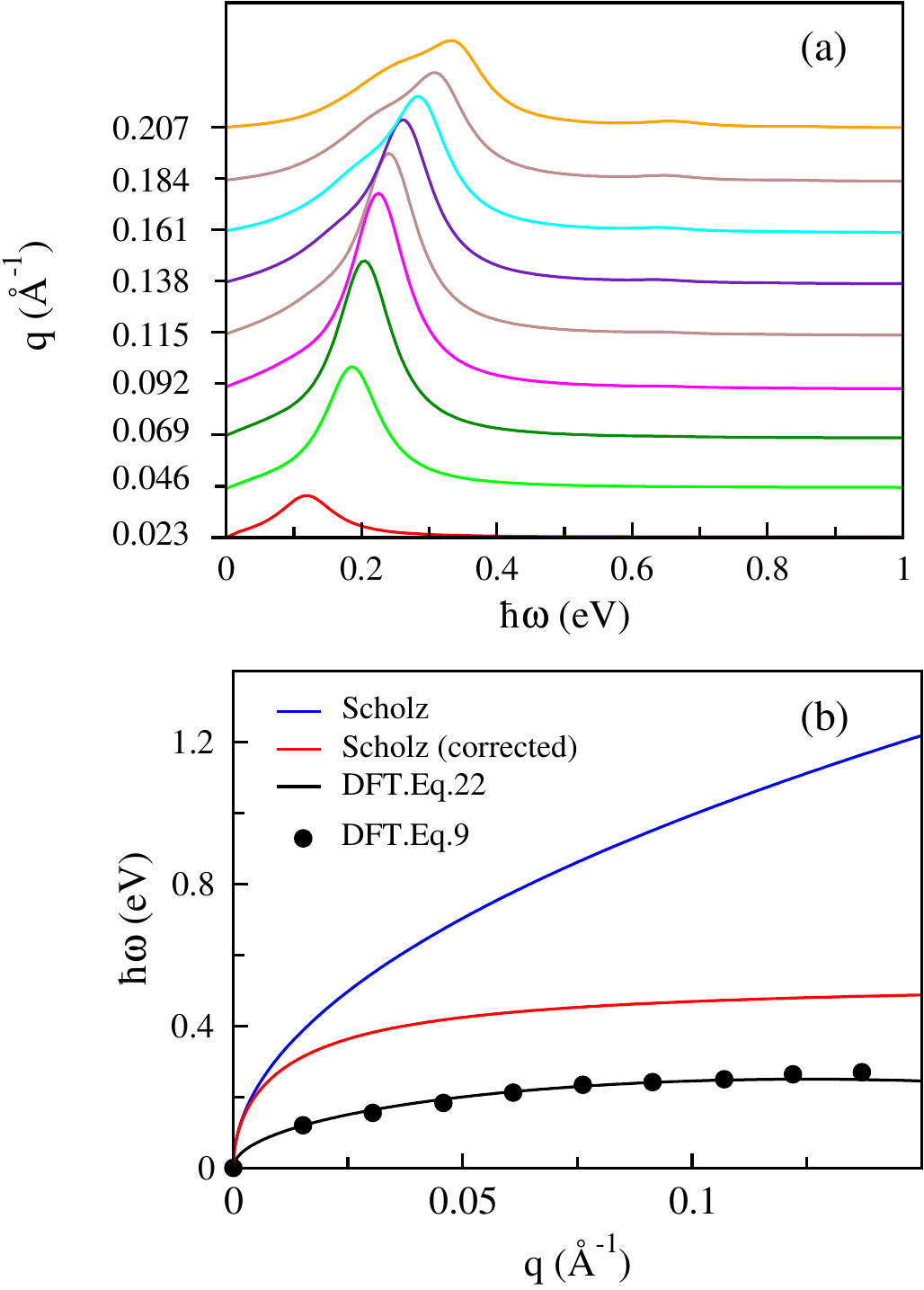}
\caption{(\textbf{a}) The EEL functions of MoS$_2$ for different $q$ and $n$$\sim$$5.6\times10^{13}$\,cm$^{-2}$ for $\epsilon=1$. (\textbf{b})~Our~optical plasmon mode of MoS$_2$ (dots), obtained from the peaks in the EEL functions in the (\textbf{a}), in comparison with a model Hamiltonian by Scholz {et al.}~\cite{scholz} (blue line). The red line is obtained using the modified bare potential given by Equation (\ref{ff}) by a model Hamiltonian. The black line shows our {ab initio} calculation with the modified bare potential
 (see the text for more details).}\label{fig9}
\end{figure}

In Figure~\ref{fig9}a, the EEL function of MoS$_2$ for different values of $q$ with electron doping of \mbox{$n=5.6\times10^{13}\,$cm$^{-2}$} is depicted and the plasmon dispersion of system plotted in Figure~\ref{fig9}b (black dots). In~the following, we compare the resulting plasmon dispersion calculated by our {ab initio} model to that obtained by a model Hamiltonian.

We use Eq.11 in the paper by Scholz, et al. \cite{scholz}, where the long wavelength behavior of the plasmon frequency of monolayer MoS$_2$ can be obtained from:
\begin{equation}\label{w0}
\omega^{0}_{q}=\sqrt{\frac{e^2q}{2\pi\epsilon_0\epsilon_r}}\sqrt{\frac{(2E_F-\Delta)[E_F(\Delta+2E_F)-\lambda^2]}{4E_F^2-\lambda^2}}\nonumber
\end{equation}

Here, $\Delta$ is the energy gap and $\lambda=80$ meV. In this case, we consider $\epsilon_r=1$ because it can be compared to our {ab initio} results.
It is obvious that the plasmon mode obtained within DFT-RPA approach for monolayer MoS$_2$ differs significantly from with that calculated using the low-energy model Hamiltonian. The similar comparison was performed for hole doping in the case of MoS$_2$ in~\cite{groenewald}, and they observed strongly reduced plasmon energies and concluded that neglecting the material-specific dielectric function $\epsilon_{\alpha \beta} (\mathbf q)$ within the minimal three-band model is a severe approximation leading to unrealistic plasmonic properties.
\par
This discrepancy could be understood by looking at both the bare potential and the Lindhard function of the MoS$_2$. In order to perceive the aforementioned discrepancy, we look at the structure of the MoS$_2$ again, which basically consists of three atomic layers. In ordinary 3D materials, the effect of lattice screening is simply a re-scaling of the interaction strength by a dielectric constant. In~quasi-2D materials, however, the interaction is modified by the polarizability of the crystalline material originated from the induced charge, $n_{ind}=-\nabla\cdot {\bf P}$, where ${\bf P}$ is the polarizability~\cite{Cudazzo}. Therefore, the Poisson equation for the potential of the external point charge takes the form:
\begin{equation}\label{vr}
\nabla^2 V({\bf r},a)=-4\pi e \delta(z)-4\pi a \nabla^2_{\rho}V({\rho},z=0,a) \delta(z)
\end{equation}
where ${\bf r}=({\rho},z)$ and ${\rho}=(x,y)$, and we use the fact that ${\bf P}=-a\nabla_{\rho}V({\rho},z=0,a) \delta(z)$. In order to solve this equation, it is convenient to take the Fourier transformation of the equation to obtain $V(q,a)$. Afterwards, by taking the inverse Fourier transformation of $V(q,a)$, we can find the effective potential in real space, which is no longer $e^2/r$, and it yields:
\begin{equation}\label{vr}
V({\bf r},a)=\frac{e^2\pi [-Y_0(r/a) + H_0(r/a)]}{2a}
\end{equation}
where the Bessel function of the second kind is defined by:
\begin{equation}
Y_n(x)=\frac{J_n(x)\cos(n x)-J_{-n}(x)}{\sin(nx)}
\end{equation}
where $J_n(x)$ is the Bessel function of the first kind. For $n$ an integer, this formula should be understood as a limit. The Struve function, $H_n(x)$, solves the inhomogeneous Bessel equation. It would be worthwhile to mention that this potential was proposed in the Keldysh model~\cite{Keldysh}, which is based on a slab of constant dielectric value, the potential between two charges in a slab of thickness $d$. In this model, $a=d\epsilon_{\parallel}/2$ where $\epsilon_{\parallel}$ is the in-plane dielectric constant of the bulk material.

Let us look at the Fourier transformation of the modified bare potential, which is given by:
\begin{equation}\label{ff}
V(q,a)=\frac{2\pi e^2}{\epsilon(q+aq^2)}=\frac{2\pi e^2}{\epsilon q} \frac{1}{\varepsilon_{lattice}(q)}
\end{equation}
where $\epsilon$ is the average dielectric constant of the environment, $a$ might depend on the thickness of the 2D crystalline material and $\varepsilon_{lattice}(q)$ plays the role as a lattice local field factor. In the case of MoS$_2$, Qiu et al.~\cite{Qiu} fitted the Keldysh model to their {ab initio} effective dielectric function at small $q$ and obtained an effective screening length of $a=35$ \AA or the slab thickness of $d=6$ \AA. Apparently, the~exact value of the thickness of monolayer MoS$_2$ is just $d\simeq6.3$ \AA. It is interesting to note that when we consider the corrected coefficient of $(1+a \mathbf q)$ with $a= 35$ \AA~in the bare Coulomb potential, we find better agreement between plasmon dispersion of the low-energy model Hamiltonian of MoS$_2$ with that obtained using the DFT-RPA approach, although it is not yet coincident with our result. This difference can be related to the non-interacting density-density response function in two methods. Notice again that the $\chi^0$ in the DFT scheme consists of many occupied bands together with the structure of the many orbitals, and as has been discussed~\cite{asgari3,asgari4}, those play important roles in the physical properties of MoS$_2$. This means that the $\chi^{0}$ calculated within DFT-RPA contains some effects of the exchange-correlation of the system. These results also show that for quasi-two-dimensional materials, the Coulomb potential should be modified properly.
At the end of this section, it is worth noting that when the potential in Equation (\ref{ff}) is used to calculate plasmon modes in MoS$_2$ using DFT simulations, we obtain very similar results, which are shown with a black line in Figure \ref{fig9}b.

\subsection{Bilayer MoS$_2$}
\begin{figure}[H]
	\centering
	\includegraphics[width=11 cm]{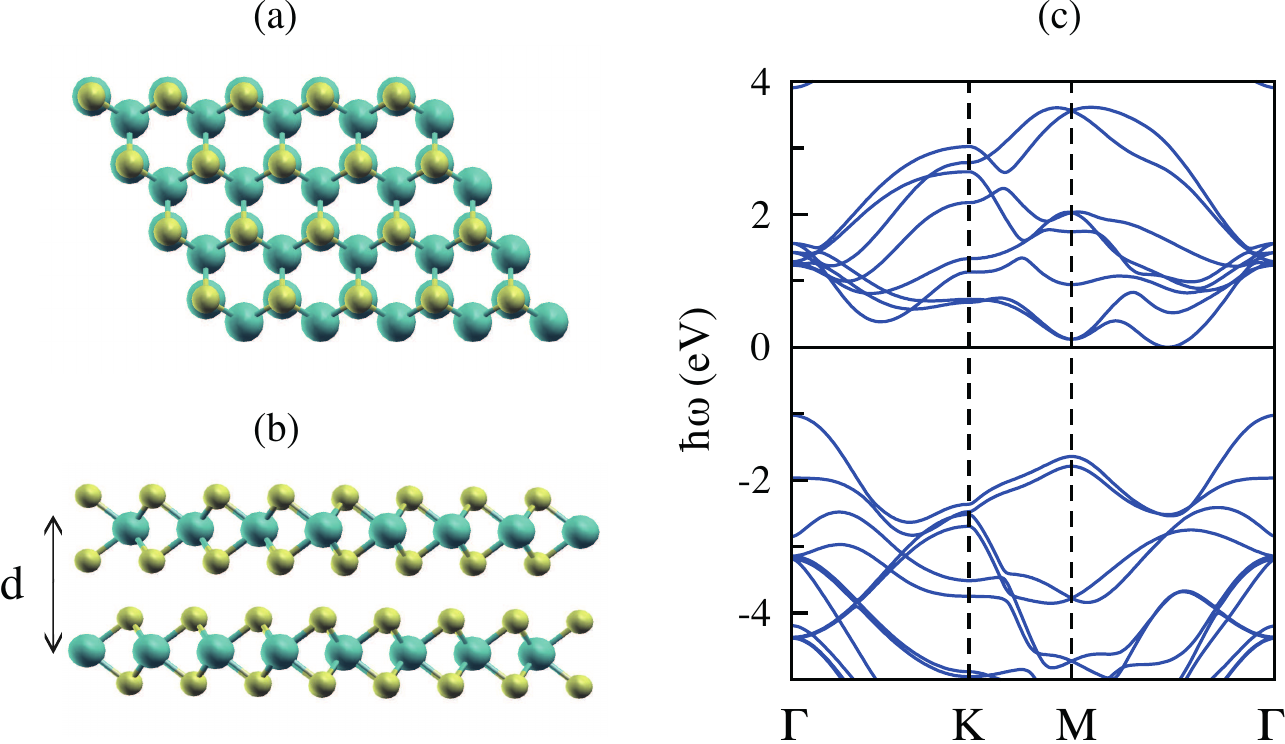}
	\caption{(\textbf{a},\textbf{b}) Top and side view of bilayer molybdenum disulfide (BL.MoS$_2$). (\textbf{c}) The band structure of BL.MoS$_2$ along the high symmetry $\Gamma- K- M- \Gamma$ directions. The Fermi level is set at 0 eV, and interlayer separation $d$ is $5.89$ \AA.}\label{fig10}
\end{figure}

The AA' stacking, in which the layers are exactly aligned, with Mo over S is the stablest stacking for the MoS$_2$ bilayer, and it is shown in Figure~\ref{fig10}a,b.
From the band structure plotted in Figure~\ref{fig10}c, based on our DFT simulations, along the high symmetry $\Gamma- M- K- \Gamma$ directions, it can be seen that bilayer MoS$_2$ has an indirect band gap in contrast to the direct band gap of the corresponding monolayer.
In fact, the conduction band minima are located between the $\Gamma$ and $K$ high symmetry points ($Q$ point), while the valence band maxima are located at the $\Gamma$ point of the BZ, revealing the indirect band gap, and this is in good agreement with previous calculations~\cite{cheiw,debbichi,komsa}.

More analysis of the band structure of MoS$_2$ can show why this material experiences an indirect to direct band gap transition when its bulk or multilayers are replaced by a monolayer. In fact, the states originating from mixing of Mo($d_{z^2}$) orbitals and the S($p_z$) orbitals at $\Gamma$ are fairly delocalized and have an antibonding nature. With increasing the separation between consecutive MoS$_2$ layers, the~layer-layer interaction decreases and lowers the energy of the antibonding states, and consequently, the valence band maximum shifts downward. The states at $K$, which are of the $d_{xy}-d_{x^2-y^2}$ character, are mostly unaffected by interlayer spacing. Thus, in the limit of widely-separated planes, i.e., monolayer MoS$_2$, the material becomes a direct gap semiconductor~\cite{ramasub}.

\begin{figure}[H]
	\centering
	\includegraphics[width=7.3 cm]{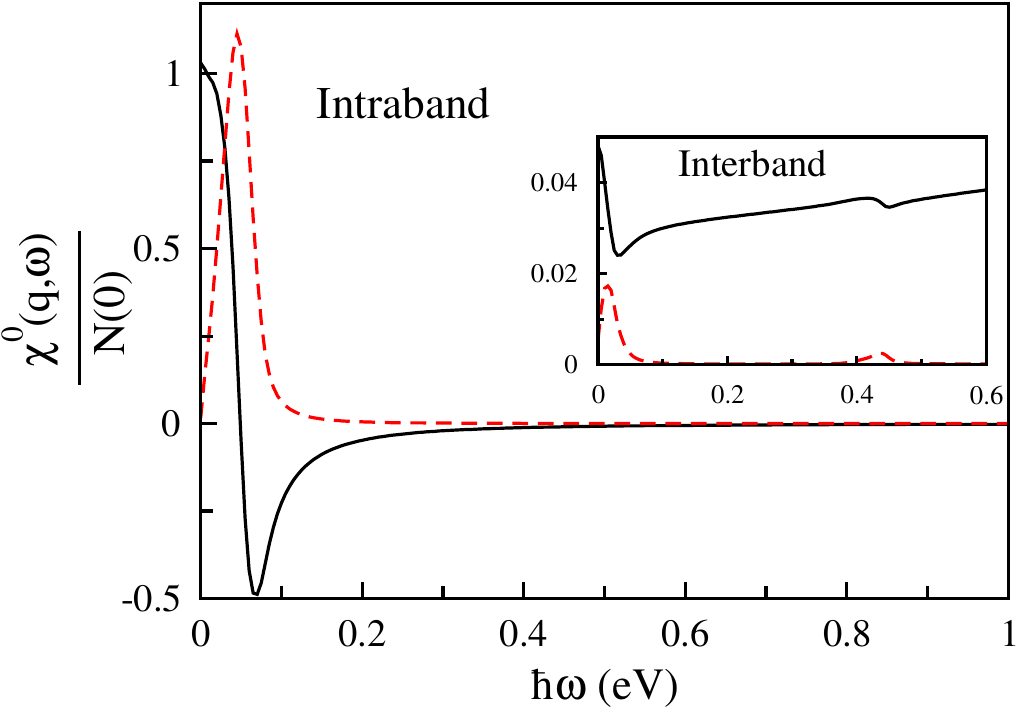}
	\caption{The non-interacting response function in units of the Fermi-level density of states as a function of $\hbar \omega$ of bilayer MoS$_2$ for $q=0.084$ \AA$^{-1}$.}\label{fig11}
\end{figure}

\textls[-15]{The non-interacting response function of bilayer MoS$_2$ for $q=0.084$ \AA$^{-1}$ with the charge carrier concentration of $n=6.9\times10^{13}$\,cm$^{-2}$ shown in Figure~\ref{fig11}.
 We can see that in bilayer MoS$_2$, both~interband and intraband contributions of non-interacting response function are considerable.~The intraband contribution of this quantity is quite strong and refers to the main collective mode, although its interband term is weak, and this can lead to new modes in the system; thus, in the following, we will explain it further.}

\begin{figure}[H]
	\centering
	\includegraphics[width=7.2 cm]{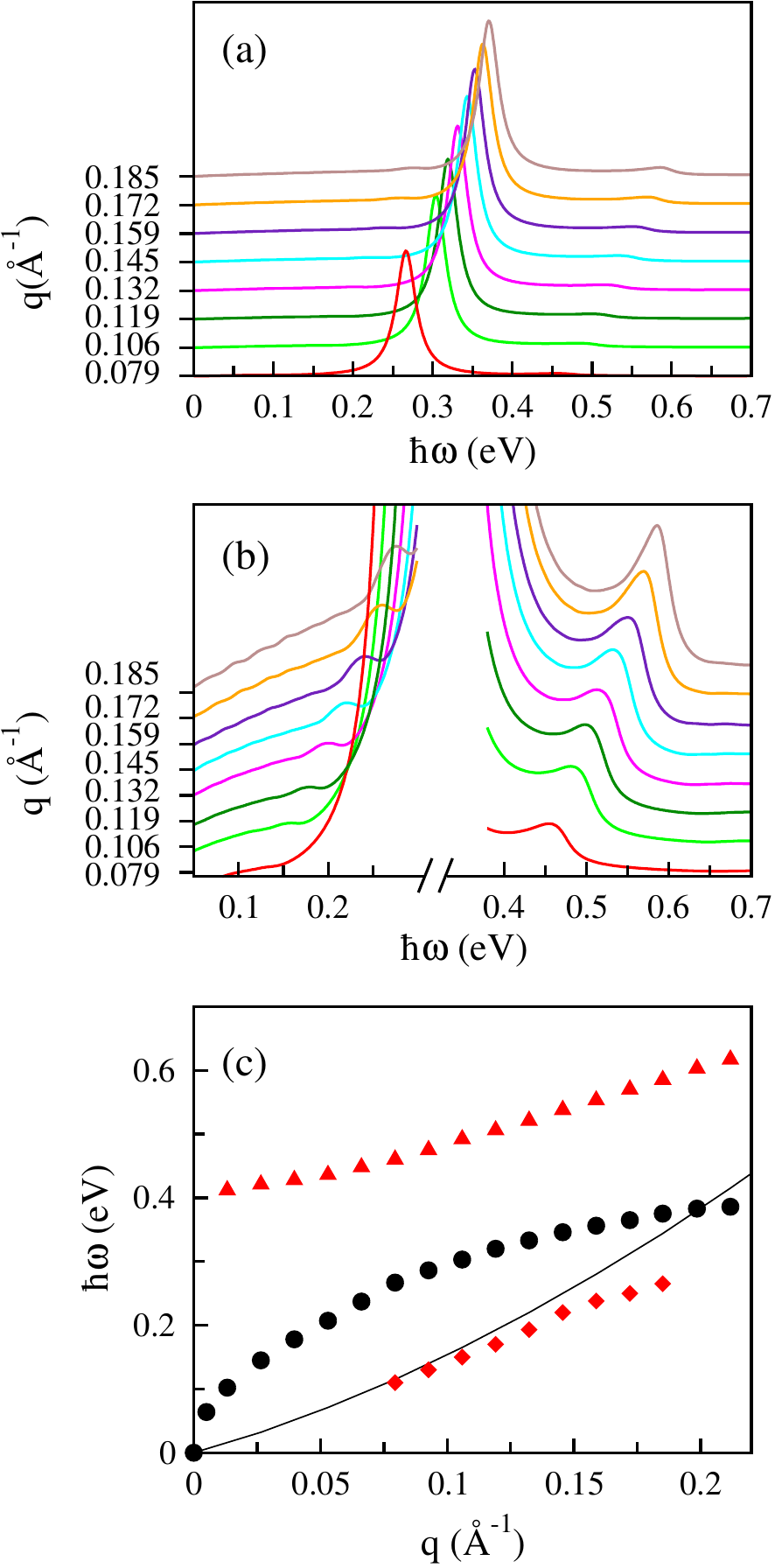}
	\caption{(\textbf{a}) The loss spectra of bilayer MoS$_2$ for various amounts of $q$ along the $\Gamma -K$ direction with $E_{\rm F}= 0.05$ eV corresponding to $n=6.9\times10^{13}$\,cm$^{-2}$. (\textbf{b}) The same as (a) for two different energy regions. (\textbf{c}) Three different plasmon modes of bilayer MoS$_2$; the black dots refer to the optical plasmon mode. The results of the acoustic and high energy modes are shown by red diamond and triangle symbols, respectively. The black line shows the boundary of the electron-hole continuum.}\label{fig12}
\end{figure}

In Figure~\ref{fig12}, we display the EEL function and plasmon spectrum of bilayer MoS$_2$ for the electron doping with $n=6.9\times10^{13}$\,cm$^{-2}$ and for energies below $1$ eV and momenta $\mathbf q$ along the high-symmetry $\Gamma-K$ direction.
We observe that plasmonic features in electron-doped bilayer MoS$_2$ are mainly characterized by a square root mode in small $q$.
In the paper by Andersen, plasmons in bilayer NbS$_2$ were studied~\cite{andersen}, and they obtained very similar plasmon dispersions.

\unskip

As previously mentioned, in the two-layer systems, we can find two plasmon modes, which can be regarded as symmetric and antisymmetric combinations of two unperturbed monolayer plasmons. One of them has a nearly linear dispersion, referred to as the acoustic plasmon mode.
\par
In this case, we find that the low-energy dynamical dielectric function in bilayer MoS$_2$ is controlled by both interband and intraband contributions, leading to an extra collective mode. Also, there are a damped high energy mode and a highly damped acoustic mode, which originates from the interband transitions. More~details of these calculations can be found in~\cite{torbatian}.

\begin{figure}[H]
	\centering
	\includegraphics[width=6.5 cm]{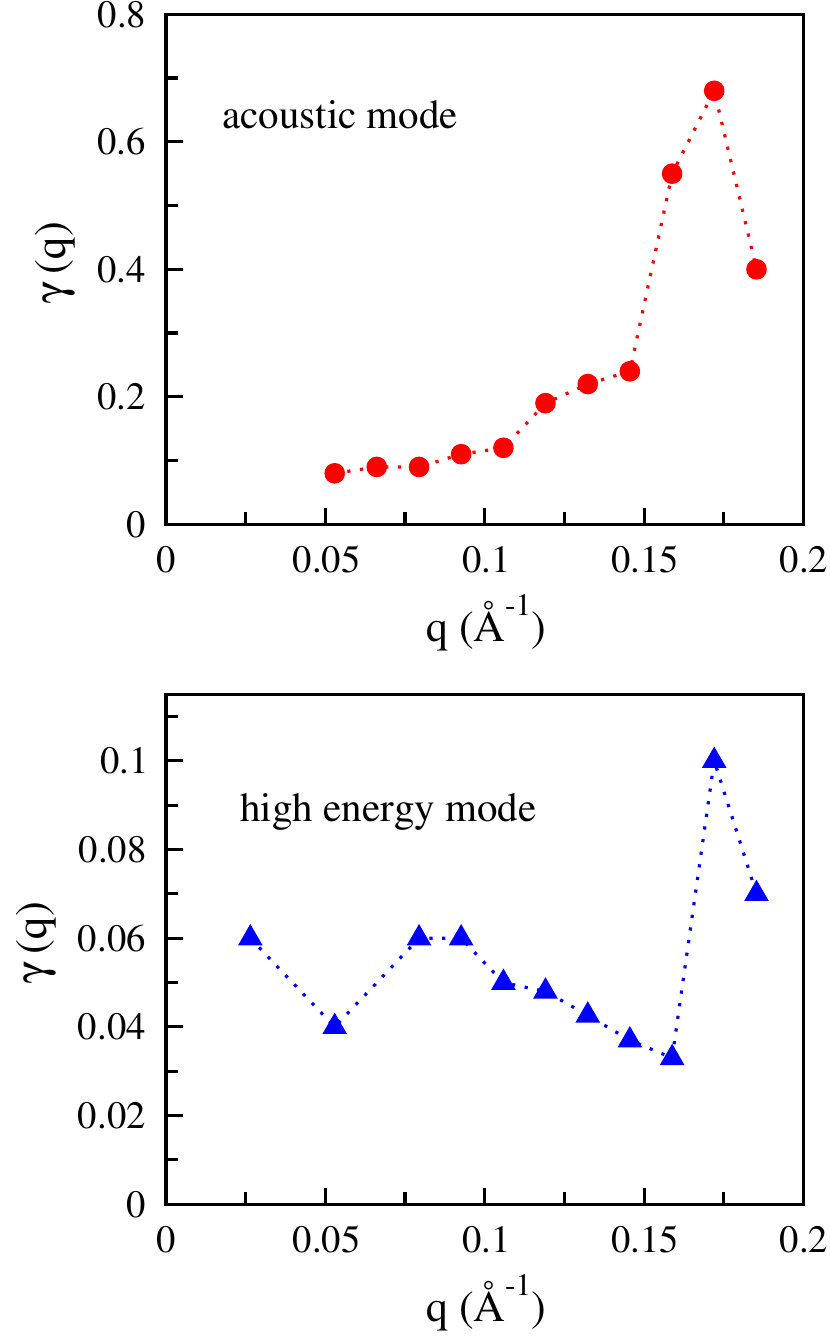}
	\caption{The damping parameter of plasmon modes in bilayer MoS$_2$ as a function of the momentum for $n=6.9\times10^{13}$\,cm$^{-2}$. Note that in order to have a well-defined plasmon mode, $\gamma(q)/\omega_p(q)<<1$. }\label{fig121}
\end{figure}

In Figure \ref{fig121}, we display the damping parameter of plasmon modes in bilayer MoS$_2$ for the electron doping value $n= 6.9 \times 10^{13}$ cm$^{-2}$ for both the acoustic and higher plasmon modes. Since the $\Im m\varepsilon(q,\omega)$ at the position of the acoustic and high plasmon modes are finite, it turns out that those collective modes are damped, and the damping parameter, $\gamma$, as a function of the momentum would be small. In this case, we should verify the condition where $\varepsilon(q,\omega_p-i\gamma)=0$ is satisfied. This equation leads to two separate equations in which the $\gamma(q)$ is given by $\gamma(q)=\Im m\chi^0(q,\omega)/\partial\Re e\chi^0(q,\omega)/\partial\omega$ at $\omega=\omega_p$ and the collective mode is obtained by $1-\nu_q\Re e\chi^0(q,\omega)-\gamma\nu_q\partial\Im m\chi^0(q,\omega)/\partial\omega=0$ again at $\omega=\omega_p$. Note that a condition required to define those plasmon modes are $\gamma(q)/\omega_p(q)<<1$. Results~shown in Figure \ref{fig121} indicate that the acoustic mode is highly damped, and the high plasmon mode is slightly damped.

\subsection{Monolayer Phosphorene}
Although two-dimensional materials such as monolayer graphene and monolayer transition metal dichalcogenide, the most common component MoS$_2$, have attracted intensive research interest owing to their fascinating electronic, mechanical, optical and thermal properties, the lack of the band gap in graphene and low carrier mobility in MoS$_2$ limits its wide applications in electronic devices and nanophotonics~\cite{schwierz,mak}.
\par
Recently, isolated two-dimensional black phosphorus (BP), known as phosphorene, with a nearly direct band gap and high carrier mobility has received enormous interest owing to its extraordinary electronic and optical properties in engineering application~\cite{zare,wang7}.
\par
Our DFT calculations for mono- and bi-layer phosphorene were carried out using the Perdew--Burke--Ernzerhof exchange-correlation functional \cite{PBE} coupled with the DFT-vdw method.

\par
\begin{figure}[H]
\centering
\includegraphics[width=8 cm]{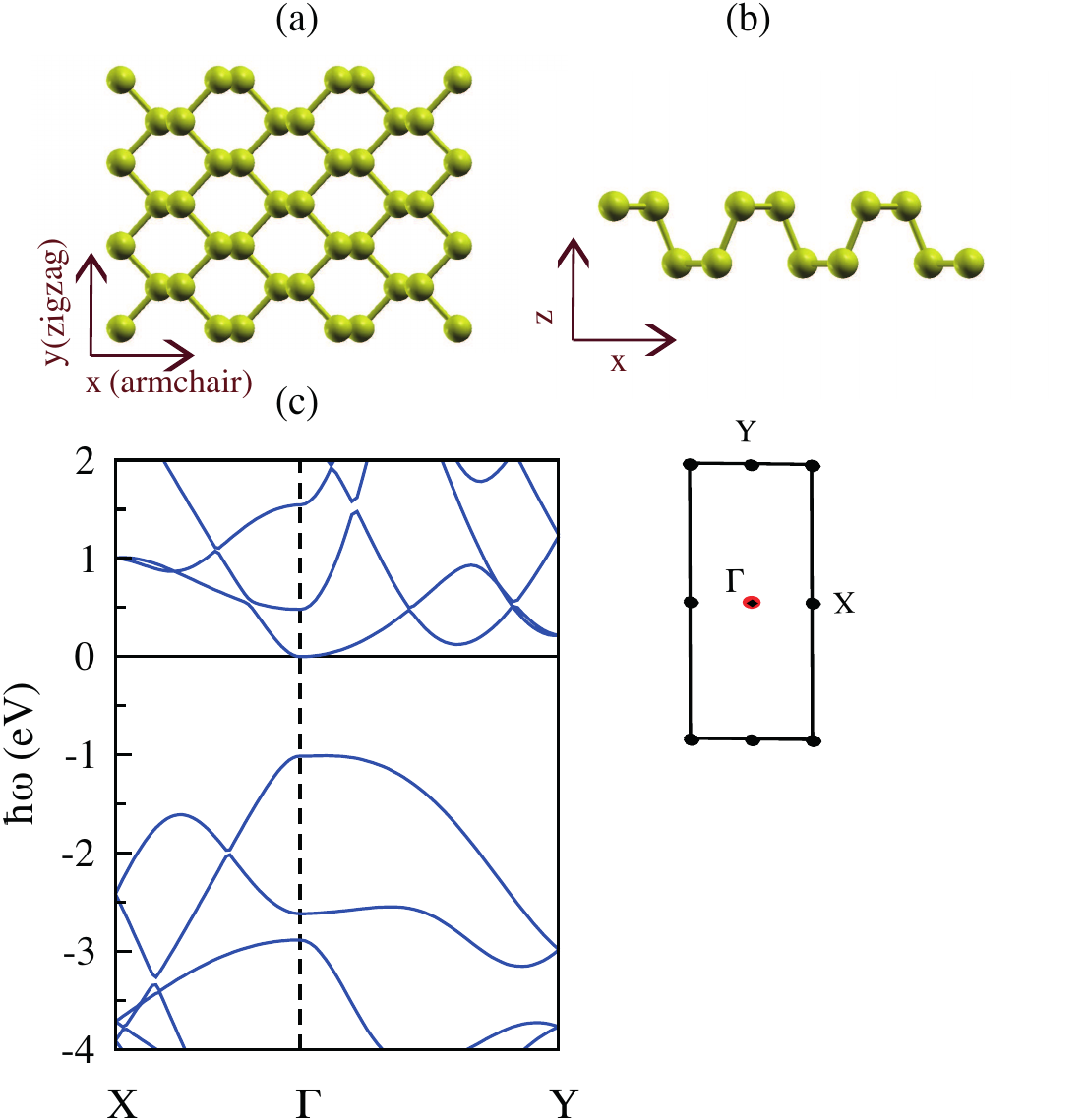}
\caption{(\textbf{a},\textbf{b}) Top and side view of the atomic structure of monolayer phosphorene. (\textbf{c}) The band structure of monolayer phosphorene along the high symmetry $X- \Gamma- Y$ directions and the associated Brillouin zone. The Fermi level is set at 0 eV.}\label{fig13}
\end{figure}

Figure \ref{fig13}a,b shows that phosphorene is a puckered honeycomb structure with each phosphorus atom covalently bonded with three neighboring atoms within a rectangular unit cell, and the two lattice constants are $a_x=4.62$ \AA~and $a_y=3.30$ \AA~along the armchair and zigzag directions, respectively~\cite{elahi}.
As a result of the puckered structure, each single-layer phosphorene contains two atomic layers where the distances between the two nearest atoms ($2.22$ \AA) and the distance between the top and bottom atoms ($2.24$ \AA) are slightly different.
\par
\textls[-10]{It should be mentioned that monolayer phosphorene is a nearly direct band gap semiconductor because the exact top of the valence band is slightly away from the $\Gamma$ point. However, they are sufficiently close to be considered as a direct band gap as previous first-principles calculations mention~\cite{tran}.}
\par
Our DFT calculations predict that the band gap of monolayer phosphorene is $0.98$ eV, and the self-energy correction enlarges the band gap to $2.0$ eV, which is ideal for potentially broad electronic applications~\cite{radis}.
The optical gap of phosphorene was reported to be $1.6$ eV by using the GW and Bethe-Salpeter equation
 method, and it is much lower than the electronic band gap of $2.15$ eV, which shows significant excitonic effects in phosphorene~\cite{ganesan}.
\par
The electronic band structure of monolayer phosphorene is plotted in Figure~\ref{fig13}c, and it shows the high anisotropic band structure with very different effective masses along the armchair and zigzag directions for both electrons and holes. This anisotropic electronic structure is useful for thermoelectric materials, because for the direction with a smaller effective mass, the carrier mobility and thus the electrical conductivity can be high, while the larger effective mass along the other direction contributes to an overall large density of states that improves the Seebeck coefficient~\cite{zare,fei}.

\begin{figure}[H]
\centering
\includegraphics[width=7.5 cm]{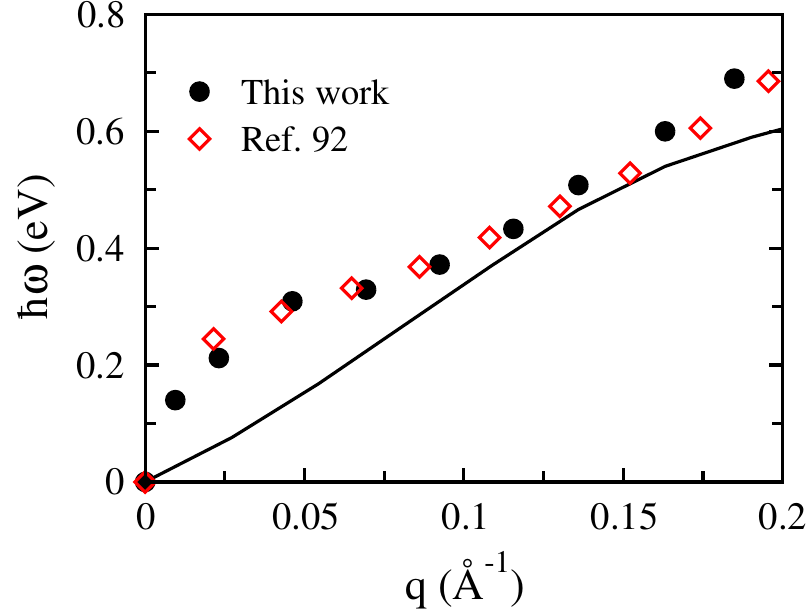}
\caption{Our calculated plasmon dispersion (dots) of monolayer phosphorene with $E_{\rm F}=0.07$ eV along the armchair direction, in comparison with
results obtained by Ghosh {et al.} for monolayer phosphorene and with the same concentration along the armchair direction (open symbols) \cite{Ghosh}.}\label{fig14}
\end{figure}

In Figure~\ref{fig14}, the plasmon spectrum of monolayer phosphorene with $E_{\rm F}=0.07$ eV along the armchair direction based on our DFT simulations is compared to the result recently reported in~\cite{Ghosh}, and they show an acceptable agreement.

\begin{figure}[H]
\centering
\includegraphics[width=7.5 cm]{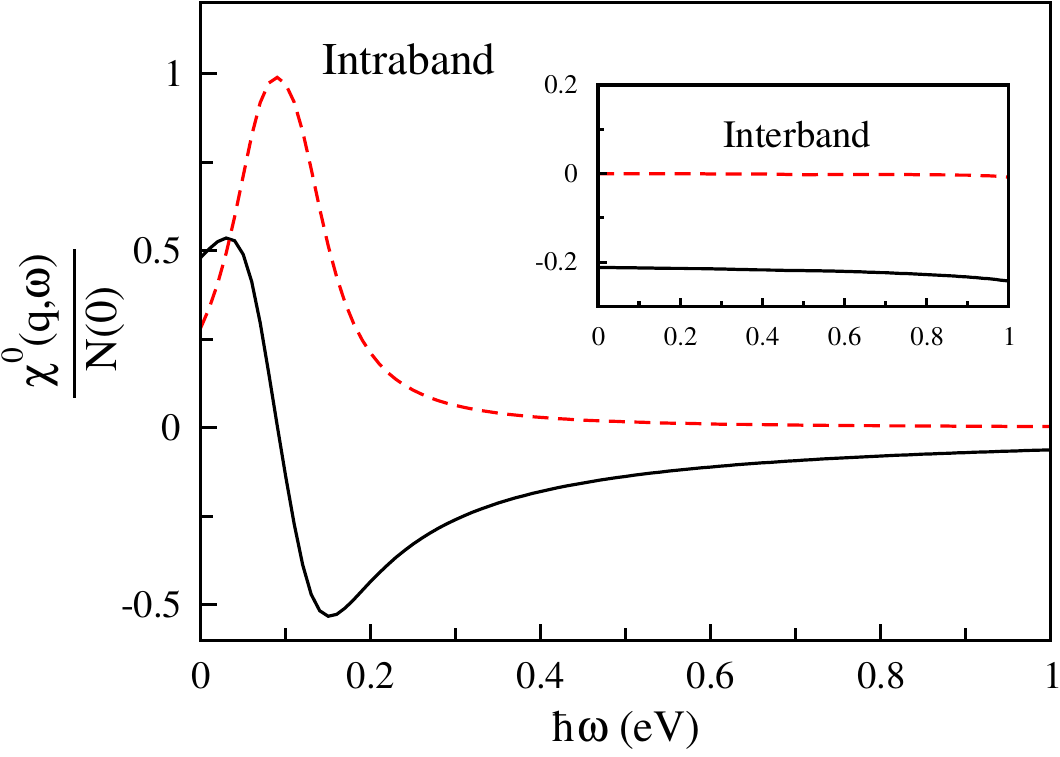}
\caption{The real (solid line) and imaginary (dashed line) parts of interband and intraband contributions of the non-interacting response function of monolayer phosphorene in units of the Fermi-level density of states as a function of $\hbar \omega$ for $q=0.046$ \AA$^{-1}$ along the armchair direction.}\label{fig15}
\end{figure}

To know the origin of the electronic excitations in the monolayer phosphorene, we obtain the different contributions (interband and intraband) of the non-interacting response function, and they are shown in Figure~\ref{fig15} for $q=0.046$ \AA$^{-1}$ and $n=3.9\times10^{13}$\,cm$^{-2}$ along the armchair direction. These~results show that the intraband term of this quantity plays an important role, and its interband term is negligible.

\par
\begin{figure}[H]
\centering
\includegraphics[width=10 cm]{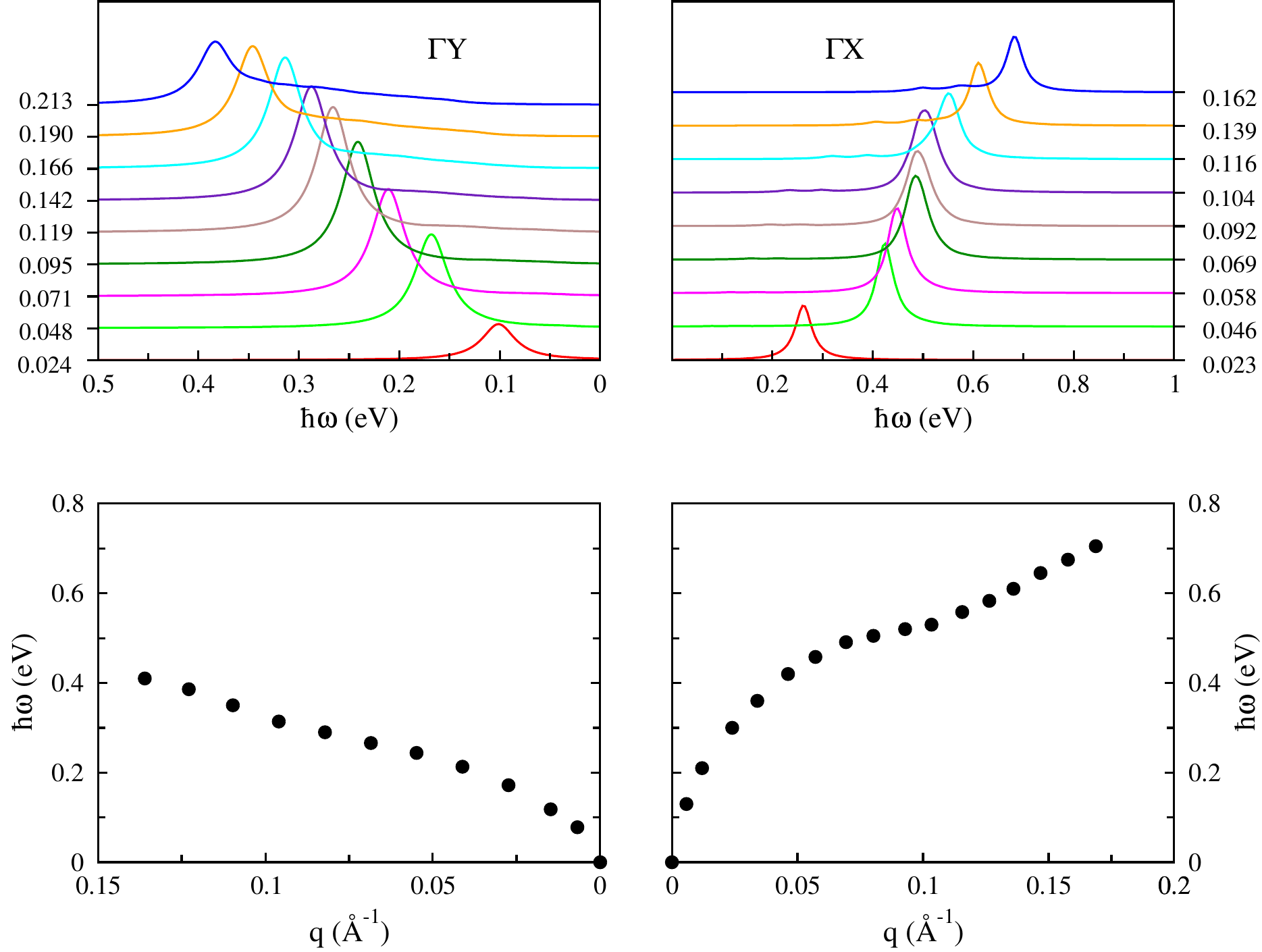}
\caption{The electron-energy loss function of monolayer phosphorene for different values of $q$ along the $\Gamma X$ and $\Gamma Y$ directions (upper panel). In the bottom panel, the plasmon modes corresponding to the peaks in the EEL function in the upper panel are plotted. In this case, the Fermi level is shifted up by $0.1$ eV, corresponding to $n=3.9\times10^{13}$\,cm$^{-2}$.}\label{fig16} 
\end{figure}

The loss spectra for different values of the $q$ and dispersion curve of the intraband plasmons of phosphorene for charge carrier concentration of $n=3.9\times10^{13}$\,cm$^{-2}$ along the armchair and zigzag directions are plotted in Figure~\ref{fig16}. The anisotropic band structure of monolayer phosphorene along the zigzag and armchair directions makes anisotropic features in the collective plasmon excitations, although they follow a low energy $\sqrt q$ dependence at the long wavelength limit. This is due to the paraboloidal band dispersion of phosphorene at low-energies, but the plasmon modes in the armchair direction have higher plasmon energy compared to the zigzag direction at the same momenta~\cite{jin,low}.

\subsection{Bilayer Phosphorene}
\begin{figure}[H]
\centering
\includegraphics[width=9.8 cm]{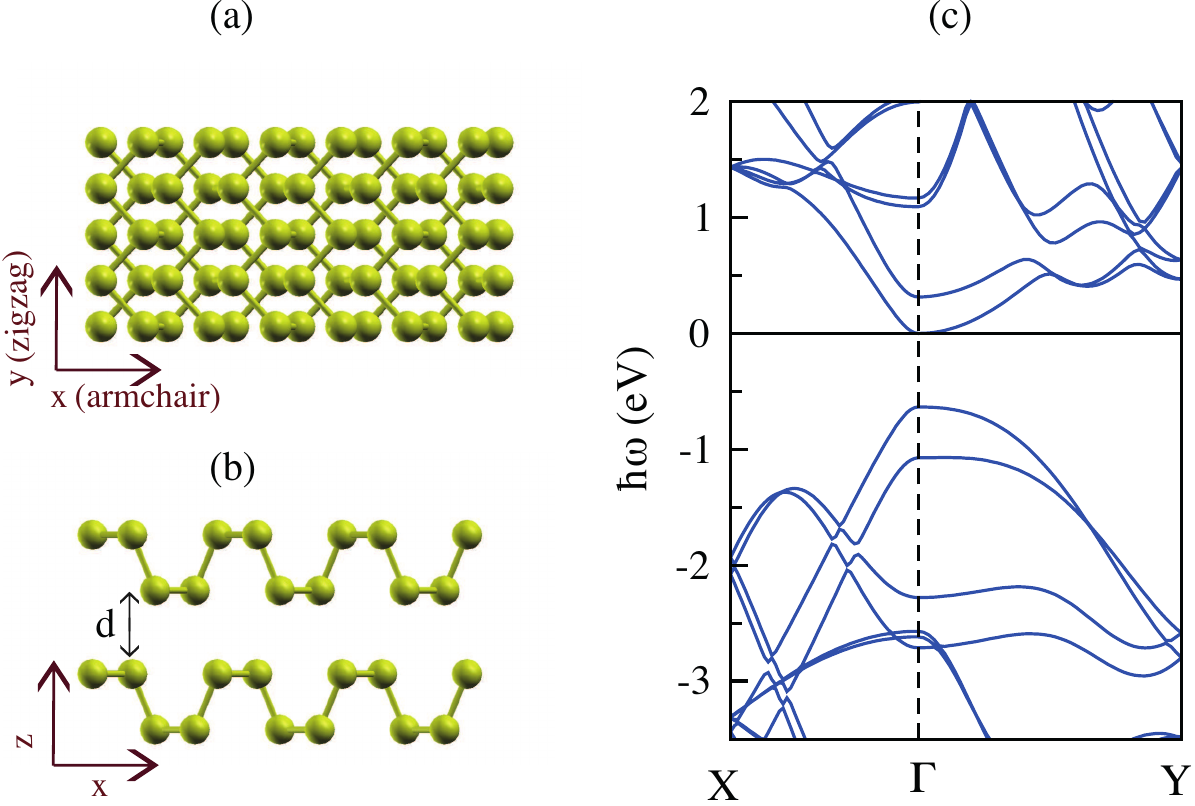}
\caption{{(\textbf{a},\textbf{b}) Top and side view of the atomic structure of bilayer phosphorene (BLP). (\textbf{c}) The band structure of bilayer phosphorene along the high symmetry $X- \Gamma- Y$ directions. The Fermi level is set at 0 eV. The optimized interlayer separation between the closest phosphorus atoms is indicated with $d$, being~$3.56$ \AA.}}\label{fig17}
\end{figure}

The stable stacking order of bilayer phosphorene is of the AB type such that the bottom layer is shifted half the lattice period along the $x$ or $y$ directions, and this is in agreement with previous studies~\cite{dai,jhun}. The crystal structure of AB-stacked bilayer phosphorene is shown in Figure~\ref{fig17}a,b.

\unskip

When two monolayers are combined to create a bilayer, the gap reduces and two additional bands emerge around the gap at the $\Gamma$ point~\cite{qiao}. Bilayer phosphorene possesses a direct band gap of $0.63$ eV, in agreement with the band gap recently reported in~\cite{jin2016}. Importantly, the inclusion of many-body effects using GW increases the band gap to $1.45$ eV.
\par
The electronic band structure of AB-stacked bilayer phosphorene is illustrated in Figure~\ref{fig17}c based on our DFT simulations.
We find that the valence band maximum is contributed by the localized states of $P$ atoms, while the conduction band minimum is partially contributed from the delocalized states, especially in the interfacial area between the top and bottom layers~\cite{caklr}.
\par
\begin{figure}[H]
\centering
\includegraphics[width=7 cm]{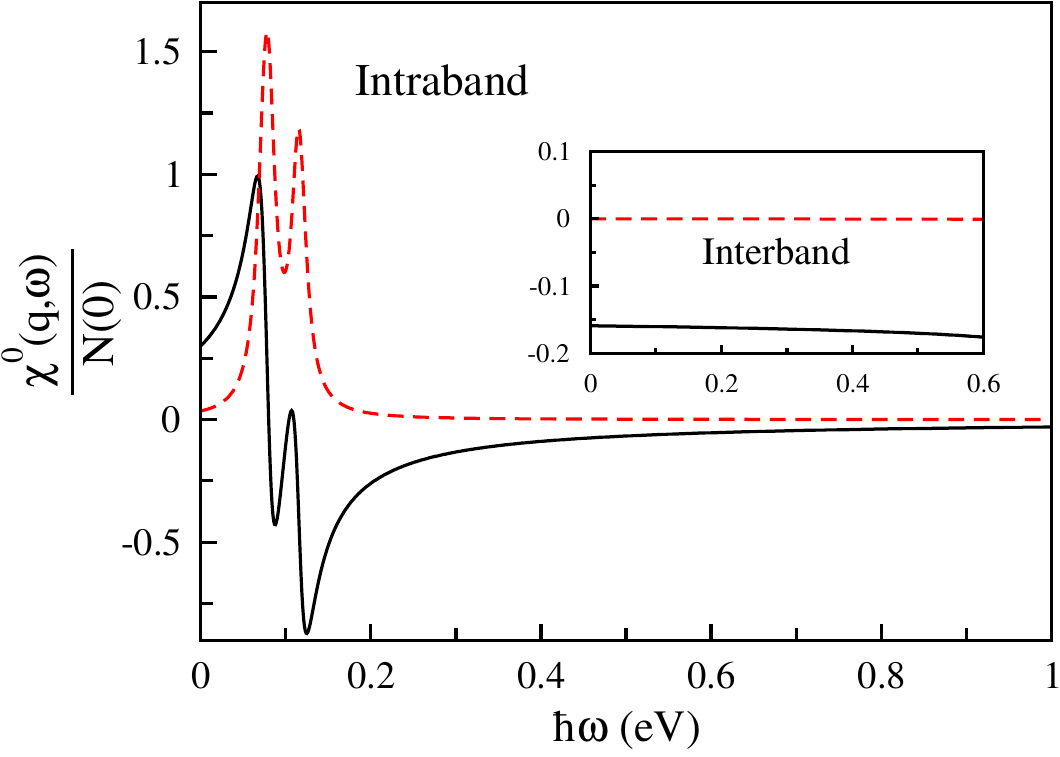}
\caption{The non-interacting response function in units of the Fermi-level density of states as a function of $\hbar \omega$ of bilayer phosphorene for $q=0.046$ \AA$^{-1}$.}\label{fig18}
\end{figure}

In order to investigate more, we illustrate the real and imaginary parts of interband and intraband contributions of the non-interacting response function of bilayer phosphorene in Figure~\ref{fig18} for a specific momentum transfer in the amount of $0.046$ \AA$^{-1}$ and $E_{\rm F}=0.05$ eV. Our {ab initio} calculations~\cite{torbatian2} show that the collective electronic excitations in bilayer phosphorene originated from intraband term of $\chi^0(q,\omega)$, and the other one is not be important enough to be considered.

\begin{figure}[H]
\centering
\includegraphics[width=6.7 cm]{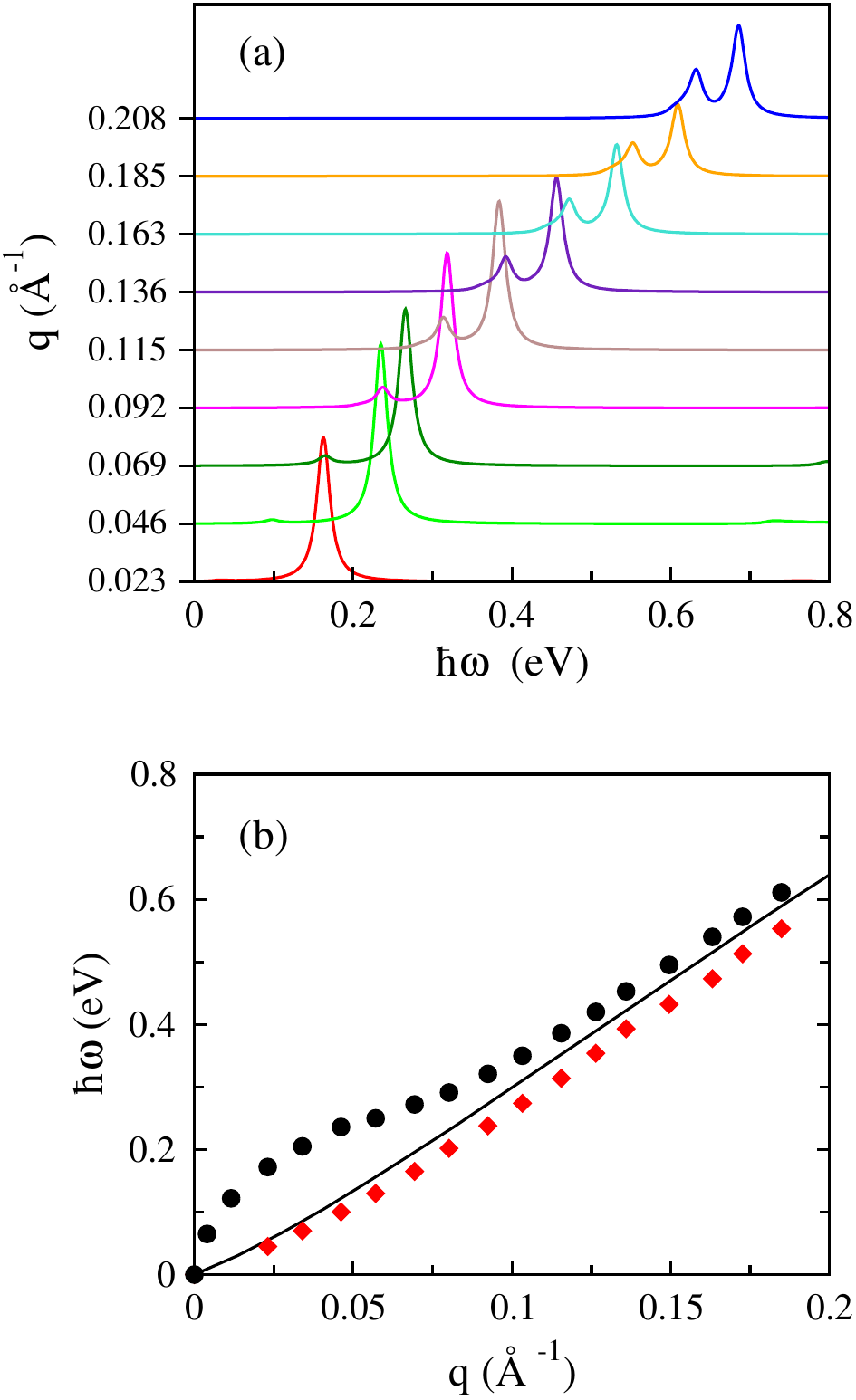}
\caption{\textls[-12]{The optical (black dots) and acoustic plasmon modes (red diamond) of bilayer phosphorene that correspond to the peaks in the EEL functions in the upper panel for $n =3.2 \times 10 ^{13}$ cm$^{-2}$ \mbox{($E_{\rm F}= 0.05$ eV)} along the armchair direction. The black line shows the boundary of the electron-hole continuum. }}\label{fig19}
\end{figure}

We illustrate the loss functions for different amounts of $q$ and also the plasmon modes of bilayer phosphorene along the armchair direction in Figure~\ref{fig19}. For this system, we have shifted the Fermi level up by $0.05$ eV to imitate the effect of finite doping. As expected, bilayer phosphorene involves two plasmon modes. One of them follows the $\sqrt q$ behavior as the optical mode, as seen in its monolayer. The other one is a damped acoustic mode with a linear dispersion at the long wavelength limit.

\section{Conclusions}

In this article, we have focused on collective modes of some pristine two-dimensional crystalline materials including graphene, MoS$_2$ and phosphorene in monolayer and bilayer structures. Our studies were based on density-functional theory simulations together with random phase approximation.
In the DFT approach, we have considered all electron band structures, and therefore, the non-interacting density-density response function could bring some effects beyond the normal Lindhard function. Furthermore, the Kohn--Sham wave functions in the vicinity of the Fermi energy are quite similar to the real and exact wave functions of the system when the system is doped. Therefore, for the doped 2D crystalline materials, the density-density response function is calculated by the Kohn--Sham wave functions and can describe the many-body effects of the system well enough to use these to explore the physical quantities of the system. In addition, the definition of the dielectric constant given by Equation~(\ref{eq11}) provides some many-body effects, which could describe why the {ab initio}-RPA calculations provide a better dispersion relation of the plasmon mode in electronic systems.
It is essential to emphasize that the material-specific dielectric function considering the multi-orbital and multiband structures (or quasi-two-dimensional polarization) such as MoS$_2$ are needed to obtain realistic plasmon dispersions. We have used a full DFT simulations together with RPA analysis to calculate the band structure, non-interacting density-density response function, the energy loss functions and, finally, plasmon dispersions of the extrinsic crystalline materials. For each material studied here, we have found different collective modes and described their physical origins. In all studied materials, the in-phase mode, which refers to the optical mode, the plasmon dispersion is displayed as a $\sqrt q$ originating from low-momentum carrier scattering. An acoustic mode on some systems is observed; however, this mode is damped. In bilayer MoS$_2$, we have observed that the plasmon modes of the electron and hole doping are not equivalent, and the discrepancy is owed to the fact that the Kohn--Sham band dispersions are not symmetric for energies above or below the zero Fermi level. The anisotropic band structure of monolayer phosphorene along the zigzag and armchair directions make anisotropic features in the collective plasmon excitations, and the plasmon mode in the armchair direction has higher energy compared to the zigzag direction at the same momenta.

\vspace{6pt}


\acknowledgments{Z.T.
 would like to thank the Iran National Science Foundation for their support through a research grant. This work is also partially supported by the Iran Science Elites Federation.}

\authorcontributions{R. A conceived and designed the paper;
Z. T and R. A analyzed the data and wrote the paper}.

\conflictsofinterest{The authors declare no conflicts of interest.}
\reftitle{References}

\end{document}